\pdfoutput=1
\documentclass[12pt,preprint]{aastex}

\defcitealias{Cern00}{CGK}

\shorttitle{$\lambda\,1.3$ and 2\,mm survey of IRC +10216}
\shortauthors{J.H. He, et al.}

\begin{document}

\title{A spectral line survey in the $\lambda\,2$ and $\lambda\,1.3$\,mm windows toward the carbon rich envelope of IRC +10216}

\author{J. H. He\altaffilmark{1,2}, Dinh-V-Trung\altaffilmark{1}, 
        S. Kwok\altaffilmark{3,1}, H. S .P. M\"{u}ller\altaffilmark{4}, Y. Zhang\altaffilmark{3},
        T. Hasegawa\altaffilmark{1}, T. C. Peng\altaffilmark{1} and Y. C. Huang\altaffilmark{1}} 




\altaffiltext{1}{Institute of Astronomy and Astrophysics, Academia Sinica, P.O. Box 23-141, Taipei 10617 Contact author: J.H. He, email: jhhe@asiaa.sinica.edu.tw}
\altaffiltext{2}{National Astronomical Observatories/Yunnan Observatory, Chinese Academy of Sciences, PO Box 110, Kunming, Yunnan Province 650011, PR China }
\altaffiltext{3}{Department of Physics, Faculty of Science, University of Hong Kong, Pokfulam Road, Hong Kong}
\altaffiltext{4}{I. Phyikalisches Institut, Universit\"{a}t zu K\"{o}ln, Z\"{u}lpicher Str. 77, 50937 K\"{o}ln, Germany}

\begin{abstract}

We present the results of our spectral line surveys in the $2$\,mm and
$1.3$\,mm windows toward the carbon rich envelope of \object{IRC
  +10216}. Totally 377 lines are detected, among which 360 lines
  are assigned to 57 known molecules (including 29 rare isotopomers
  and 2 cyclic isomers). Only 17 weak lines remain unidentified.
Rotational lines of isotopomers $^{13}$CCH and HN$^{13}$C are
detected for the first time in \object{IRC +10216}. The detection
  of the formaldehyde lines in this star is also confirmed. Possible
abundance difference among the three $^{13}$C substituted
isotopic isomers of HC$_3$N is reported. Isotopic ratios of C and O are
  confirmed to be non-solar while those of S and Si to be nearly
  solar. Column densities have been estimated for 15 molecular
species. Modified spectroscopic parameters have been calculated for
NaCN, Na$^{13}$CN, KCN and SiC$_2$. Transition frequencies from the
present observations were used to improve the spectroscopic parameters of Si$^{13}$CC, $^{29}$SiC$_2$ and $^{30}$SiC$_2$. 

\end{abstract}


\section{Introduction}

\object{IRC +10216} (CW Leo) is a nearby AGB star with heavy mass loss
of about $3\times 10^{-5}$\,M$_\odot$\,yr$^{-1}$. The distance to \object{IRC
  +10216} is estimated to be in the range of 120-150\,pc
\citep{Groe98,Luca99}. The combination of proximity and the massive
expanding envelope created by the high mass loss rate makes \object{IRC+10216} one of the strongest sources of continuum and molecular line emission in the radio sky. Over 60 molecular species have been identified in \object{IRC +10216}, some of which are unique to this source. The rich molecular inventory is the result of active photochemistry and molecule-radical reactions initiated by the penetration of interstellar UV photons into the outer envelope of \object{IRC +10216} \citep{Cher93,Mill94,Mill00}. 

Due to the expansion of the envelope, the physical conditions (i.e.,
temperature, gas density, and infrared radiation field) vary greatly
within the envelope of \object{IRC +10216}. Such condition allows the
excitation of many different molecular species by collisions with
hydrogen molecules. We expect that line emission from not too heavy
molecules such as diatomic or triatomic molecules and simple carbon chains to be strong in the millimeter region.

Several spectral line surveys have been carried out toward \object{IRC +10216}. The most comprehensive and most sensitive surveys to date are those of \cite*{Cern00} (\citetalias{Cern00} survey) in the 2\,mm window ($129-172.5$\,GHz) with the IRAM 30\,M telescope and \cite{Kawa95} in the $30-50$\,GHz range with Nobeyama 45\,M telescope. In other wavelength regions, the other surveys (\citealp{Aver92} and \citealp{Groe94} in the
230 and 345\,GHz bands, respectively, \citealp{Joha85} in the
$70-115$\,GHz band) have lower sensitivity.

The advent of millimeter and sub-millimeter radio interferometers with
wide frequency coverage such as the SMA (Sub-Millimeter Array), CARMA
and PdBI highlights the need for sensitive spectral line survey in
these wavelength regions to serve as a guide for future interferometric observations of circumstellar envelopes. In this paper we present sensitive spectral line surveys toward \object{IRC +10216} in the 2\,mm ($130-160$\,GHz) and 1.3\,mm ($219.5-267.5$\,GHz) regions using the radio telescopes of the Arizona Radio Observatory (ARO).  

\section{Observations and data reduction}

The surveys were carried out using the $\lambda\,2$\,mm receiver of
the Arizona Radio Observatory (ARO) 12M Telescope (KP12M) at Kitt Peak
(April 26, 2005 to April 29, 2006) and the $\lambda\,1.3$\,mm JT
receiver of the Heinrich Hertz submillimeter Telescope (SMT) at Mt. Graham (October 31, 2005 to January 24, 2007). Observed
frequency ranges are 131.2-160.3\,GHz for the $\lambda\,2$\,mm survey,
219.5-245.5 and 251.5-267.5\,GHz for the $\lambda\,1.3$\,mm
survey. The sky subtraction was made with beam switch with a beam
throw of 2 arcmin in azimuth at a rate of 1.25\,Hz (KP12M)
  or 1\,Hz (SMT). The receivers were configured in single
  sideband (SSB) dual
  polarization mode. At
  the KP12M telescope, the image sideband signal was rejected by tuning
  the backshots, while for the SMT telescope, Martin-Pupplet filter was used
  to reject the image sideband. The image rejection ratio was usually
better than 20\,dB.  Millimeter Auto-Correlators (MAC, 3072
channels, 0.195\,MHz per channel) were used for the $\lambda\,2$\,mm
survey, while Forbes Filterbanks (FFBS, 1024 channels, 1\,MHz per
channel) were used for the $\lambda\,1.3$\,mm survey. The antenna
temperature scale was determined by vane calibration. Typical system
temperature $T_{\rm sys}$ was 250\,K for the $\lambda\,2$\,mm
data and 400\,K for the $\lambda\,1.3$\,mm data (below 300\,K and 500\,K for $44\%$ and $88\%$ of all data,
  respectively). An integration
of one hour resulted in a typical sensitivity in $T^*_{\rm R}$ or
  $T^*_{\rm A}$ of about 10-20\,mK for a resolution of 1\,MHz in both
  surveys. The atmospheric opacity was better than 0.2 and 0.4 for
  $60\%$ and $90\%$ of our data (in both $1.3$\,mm and $2$\,mm ranges), 
respectively. Both repeated observations by engineer and the
  analysis of the $T_{\rm sys}-e^{-\tau_0 A}$ relation of our data
  ($\tau_0$ is zenith atmospheric opacity, $A$ is airmass)
  demonstrate a relative calibration uncertainty (repeatability)
  within $10\%$ (adopted in the determination of excitation
  temperatures and isotopic ratios). However, the absolute calibration error could be
  higher than this (we adopt $20\%$ for the determination of column density). 

The data were reduced using the Gildas software package CLASS. Bad
scans were discarded and bad channels were removed. Baseline was removed 
by fitting low order (typically 1-2 order) of polynomial
  to it. With the fixed LO and IF (${\rm IF}=5$ and $1.5$\,GHz for the
  1.3\,mm and 2\,mm surveys respectively), we identified several
  mirror line features of strong lines from the image sideband that
  leaked through the image rejection filters, using our
spectra observed 10 or 3 GHz apart. For several
scans near the higher and lower ends of our surveyed frequency ranges, strong lines found by other surveys \citep{Cern00,Aver92} were used to check possible image sideband contamination. The image sideband contamination features have been marked in the spectral plots in Fig~\ref{spplot}.

The $\lambda\,2$\,mm survey data were recorded in $T^*_{\rm R}$, the
antenna temperature corrected for atmospheric attenuation, Ohmic loss
and rearward and forward scattering and spillover. A corrected
mainbeam efficiency of $\eta^*_m = 0.75$ (from the KP12M telescope
manual) was used to derive the mainbeam temperature through $T_{\rm
  mb} = T^*_R/\eta^*_{\rm m}$. The FWHM beam size of the KP12M is
$43\arcsec$ at 145\,GHz.

The $\lambda\,1.3$\,mm survey data were recorded in $T^*_{\rm A}$, the
antenna temperature corrected for atmospheric attenuation, Ohmic loss
and rearward scattering and spillover. A special main beam efficiency
of $\eta_{\rm m}$ was used to calculate the mainbeam temperature
through $T_{\rm mb} = T^*_{\rm A}/\eta_{\rm m}$, where $\eta_{\rm m} =
0.65$ for polarization channel A and 0.55 for channel B (private
  communication with Dr. William Peters at ARO). The FWHM beam size of
the SMT is $31\arcsec$ at 240\,GHz.

\section{Line identification}
\label{lineiden}

Line identification was performed by using Cologne Database for
Molecular Spectroscopy \citep[CDMS, see][]{Muel01,Muel05}\footnote{The
  CDMS catalogue at \url{http://www.ph1.uni-koeln.de/vorhersagen/}} and
the Molecular Spectroscopy database of Jet Propulsion Laboratory
\citep[JPL, see][]{Pick98}\footnote{The JPL catalogue at
  \url{http://spec.jpl.nasa.gov/home.html}}. The online Lovas line
list \citep{Lova04}\footnote{The Lovas line list at
  \url{http://physics.nist.gov/PhysRefData/Micro/Html/contents.html}}
was also frequently consulted to help verify the line identities. We
first identified clearly-detected lines. Weaker lines of
the identified species (particularly those asymmetric rotors) are then
checked by comparing the predicted LTE line intensities
from CDMS or JPL catalogues with the spectral data. Especially the line
frequency predictions of SiCC isotopologues computed during
this work (see details below) helped us identify many unidentified or
weak lines. To implement the CDMS and JPL catalogues, we developed
several scripts on the basis of CLASS to automate the line
identification. Our scripts also invoke the truncated parabolic profile fitting procedure of CLASS to fit the line profiles with an truncated parabolic profile to determine line center frequency and line center antenna temperature. Integrated intensities of blended lines were determined by the profile fitting as well, while those of isolated lines were obtained by direct integration of the observed line profiles.

In the course of the present investigation, entries for a number of molecular species were created for 
the CDMS as comprehensive predictions were missing in either the CDMS or JPL catalogues. The species are 
NaCN/NaNC, Na$^{13}$CN/NaN$^{13}$C, KCN/KNC, Si$^{13}$CC, $^{29}$SiC$_2$, and $^{30}$SiC$_2$. 
In addition, an entry was also provided for the main isotopic species $^{28}$SiC$_2$.
The alkali metal cyanides are nearly T-shaped molecules in the gas phase with the M-C bond slightly longer 
than the M-N bond \citep{NaCNrot,KCNrot3}. Therefore, these species should be considered more as 
alkali metal isocyanides. However, following the current convention,
we still use the designation of cyanides. 

NaCN \citep{NaCNrot} and KCN \citep{KCNrot1,KCNrot2,KCNrot3} have been characterized by microwave spectroscopy; 
the latter also by millimeter-wave spectroscopy. Ab initio calculations (H.~S.~P. M\"uller, unpublished) 
yield very large $a$-dipole moment components of about 8.85 and 9.96~D for NaCN and KCN, respectively, 
while the $b$-components are only of order of 0.2~D. Laboratory investigations employing molecular beams 
\citep{NaCNrot,KCNrot3} permitted $b$-type transitions to be observed despite the small dipole 
moment component for each molecule. These transitions ensure reasonably good predictions into the 
millimeter-wave region. Predictions in the upper millimeter-wave region or at higher values of $K_a$ 
should be viewed with caution because of the fairly small datasets, especially in the cases of NaCN and 
Na$^{13}$CN. Earlier fit of the rotational spectra of these molecules only involved four quartic 
centrifugal distortion terms (planar reduction). New fits with five
such terms from Watson's S reduction were performed in this work, which was an obvious choice for these
asymmetric molecules very close to the prolate limit. The
inclusion of NaCN transition frequencies determined through radioastronomical observations 
have negligible effects on the spectroscopic parameters and their uncertainties. These frequencies 
were thus omitted from the final fits. The fitted results are given in Table~\ref{MCN-parameters}. 
Predictions of the rotational spectra as well as input files to generate these are available in the CDMS. 

The rotational spectrum of the main isotopic species of SiC$_2$ had been studied by 
\cite{SiC2-1-0} and \cite{SiC2-rot}. The derived spectroscopic parameters yield sufficiently good predictions for 
the transitions observed toward \object{IRC +10216} thus far. However, at higher frequencies the predictions become 
imprecise. The discrepancies between the earlier predictions and observed frequencies reached four times the quoted uncertainties, even though the number of parameters 
(15) was comparatively large with respect to the number of transition frequencies (34). 
Trial fits with different sets of spectroscopic parameters showed that a significant improvement was 
possible if Watson's $S$ reduction was used instead of the $A$ reduction. But even with two more 
parameters used in the fit of now 35 lines, the transition frequencies from \citet{SiC2-rot} could be reproduced to slightly worse than the quoted uncertainties. Therefore, these uncertainties have been increased by 50\,\%.
The high order term $P_{KKKJ}$ was estimated from the lower order terms $L_{KKJ}$, $H_{KJ}$, and $D_{JK}$. 
This parameter set was the starting point for the other isotopic species. 
The initial data set for Si$^{13}$CC consisted of transition frequencies determined in the laboratory around 
350~GHz as well as frequencies from astronomical observations at lower frequencies \citep{SiC2-isos}. 
Some Si$^{13}$CC parameters were estimated from those of SiC$_2$. 
In the cases of $^{29}$SiC$_2$ and $^{30}$SiC$_2$, transition frequencies were available almost exclusively 
from astronomical observations \citep{SiC2-isos}; the $1_{01} - 0_{00}$ transitions had been measured in the 
laboratory \citep{SiC2-1-0}. Several spectroscopic parameters were kept fixed to those estimated from 
SiC$_2$, because the input data sets are rather small. The $A - (B + C)/2$ were determined by assuming 
substitutions of the Si atom do not change $A$, an assumption that usually holds well under similar 
circumstances. In the final fits for Si$^{13}$CC, $^{29}$SiC$_2$, and $^{30}$SiC$_2$, transition frequencies 
from the current study were also used. The final sets of spectroscopic parameters have been gathered in 
Table~\ref{SiC2-parameters}. Predictions of the rotational spectra as well as input files to generate 
these are available in the CDMS. Because of the limited number of transition frequencies used in the fits 
all predictions should be taken with great caution.

\section{Results}
\label{results}

 A total of 377 line features are detected in our surveys. Among
  them, 360 are assigned to 57 known molecules (including 29 rare
  isotopomers and 2 cyclic isomers). Only 17 weak lines remain
  unidentified and need further observations to confirm. The two rotational lines from formaldehyde (H$_2$CO) that were first identified by
  \cite{Ford04} in this carbon star are confirmed by our
  observation. The fact that the $2_{1,2}-1_{1,1}$ (140840\,MHz)
  line is the strongest line of H$_2$CO indicates that its excitation
  temperature 
  could be lower than 10\,K.

\begin{figure*}[tbH]
\centering
\begin{minipage}[t]{0.6\textwidth}
\includegraphics[angle=270,scale=.5]{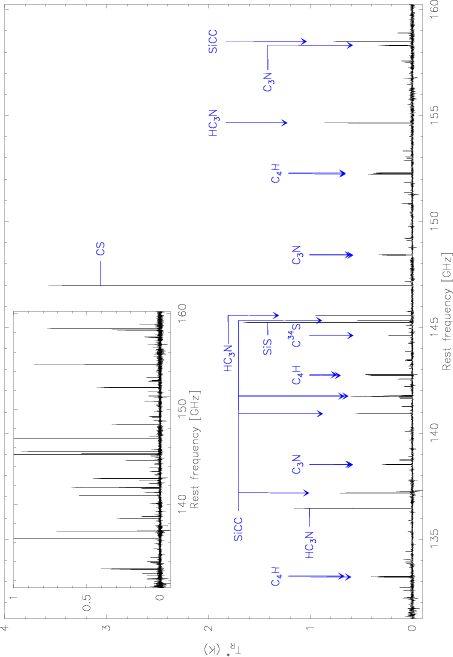}
\end{minipage}\\
\begin{minipage}[t]{0.6\textwidth}
\includegraphics[angle=270,scale=.5]{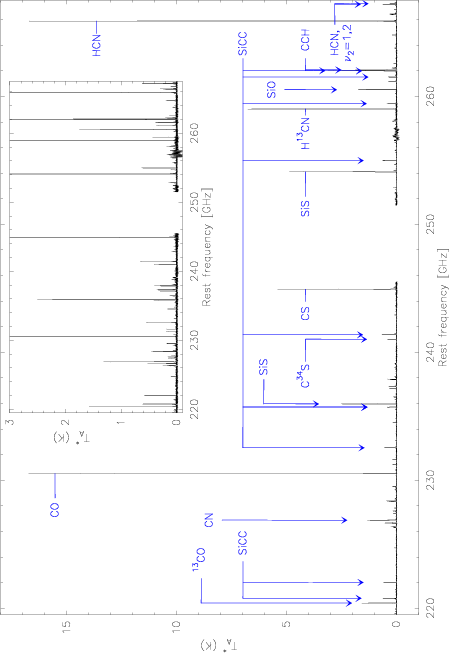}
\end{minipage}
\caption{Over view of strong lines detected in the $\lambda\,1.3$ and 2\,mm surveys. \label{ovplot}}
\end{figure*}

Overview plots of the $\lambda\,1.3$\,mm and $\lambda\,2$\,mm spectral
survey results are shown separately in the two panels of Fig.~\ref{ovplot}, with all
spectra smoothed to a resolution of about $3$\,MHz. Line features with
main beam temperature $T_{\rm mb} > 0.5$\,K in the $\lambda\,1.3$\,mm
survey and $T_{\rm mb} > 0.3$\,K in the $\lambda\,2$\,mm survey have
been labeled. Weaker features are enlarged in the insets of each
panel. The $\lambda\,1.3$\,mm survey results are dominated by strong
lines from 11 species: CO, HCN and H$^{13}$CN with $T_{\rm mb} > 10$\,K; CS, SiS, SiO, CCH, CN and $^{13}$CO with $T_{\rm mb} > 1$\,K; SiCC and C$^{34}$S with $T_{\rm mb} > 0.5$\,K. The $\lambda\,2$\,mm survey results are dominated by strong lines from 7 species: CS, SiS and HC$_3$N with $T_{\rm mb} > 1$\,K; SiCC and C$_4$H with $T_{\rm mb} > 0.5$\,K; C$_3$N and C$^{34}$S with $T_{\rm mb} > 0.3$\,K.

 The overview plot of the $\lambda\,2$\,mm survey in Fig.~\ref{ovplot}
resembles that of \citetalias{Cern00} survey, except C$^{34}$S. But
the relative intensity of lines are significantly different. The
integrated line intensity ratios between the \citetalias{Cern00} and
our surveys for the strong lines labeled in Fig.~\ref{ovplot} have
been roughly estimated. All these ratios are larger than 1 because our
$\lambda\,2$\,mm survey has larger beam size and so weaker line
intensities due to beam dilution. Although the CS, HC$_3$N and SiCC
line intensity ratios are all roughly equal to 3, the average ratios
 of C$_4$H and C$_3$N line ratios are obviously smaller (roughly 1.8
 and 2.6, respectively), while that of SiS and C$^{34}$S are obviously
 larger (around 5.8 and 3.7, respectively). The smaller line ratios
of C$_4$H and C$_3$N indicate that their spatial distribution might be
more extended than the other species. Although the hollow shell
distribution detected by interferometry for C$_4$H \citep{Guel93} and
C$_3$N \citep{Bieg93} showed similar radius of about $15\arcsec$ as
other extended species such as SiCC \citep{Luca95} and HC$_3$N
\citep{Bieg93}, the smaller line intensity ratios of C$_4$H and C$_3$N
might indicate additional extended emission component not detected by
interferometry observations. The larger line intensity ratios of SiS and C$^{34}$S indicate
that their spatial distribution might be more compact than the other species. This result of SiS agrees with the interferometry observation by \cite{Luca95} in which they found a strongly centrally peaked distribution of SiS line emission. The slightly higher compactness of C$^{34}$S can be interpreted by its centrally peaked spatial distribution and much lower abundance than CS.

The full resolution spectra of both the $\lambda\,1.3$\,mm and 2\,mm surveys are present in Fig.~\ref{spplot}.  The shown temperature is $T^*_{\rm R}$ for the 131.2-160.3\,GHz range and $T^*_{\rm A}$ for the 219.5-245.5 and 251.5-267.5\,GHz ranges. All spectra in Fig.~\ref{spplot} have been smoothed to roughly 1\,MHz resolution by box averaging. Temperature is plotted usually in -0.1 to 0.3\,K range, with exceptional plots marked by a solid square to the upper-right corner. Vertical arrows are used to mark the catalogue frequencies. The Quantum number (Qn) convention conforms to that of the CDMS and JPL catalogues. The involved Qns are: primary Qns N, K, Ka, Kc, $\Lambda$, $\ell$, v, and $\nu_n$; spin involved Qns J and F; J$_{\rm K}$ for symmetric tops; J$_{\rm Ka,Kc}$ or in rare cases N$_{\rm Ka,Kc}$ for asymmetric tops; $\nu_n^\ell$ for the excited bending mode $\nu_n$ of a linear molecule with its projection $\ell$ along the molecular axis; e/f for the parity of upper level, according to the convention of \cite{Brow75}. For the special case of {\it l}-C$_3$H and its isotopic species, the projection of electronic orbital angular momentum on the molecular axis has a quantum number $\Lambda=1$. So its ground level is X$^2 \Pi$. In the excited bending mode ($\nu_4=1$, with a projection quantum number $\ell=+1$ on the molecular axis), only the lower component of the $\Lambda$ doublets is detected and this component is denoted as $\mu^2 \Sigma$ in which $\mu$ means the lower component. An online color version of the paper is available in which blue, magenta and red labels are used to differentiate identified, unidentified and image sideband contamination lines respectively, and zoom-in plots are shown in red boxes. 

All identified species are summarized in Table~\ref{moltab} which
gives the number of detected transitions and the index of the tables
(Table~\ref{tab6}~-~\ref{tab11}) that contain the relevant line
parameters. Table~\ref{tab4} gives line parameters of all detected lines (with its format explained below). The identified lines are divided into 6 groups with their line parameters extracted from Table~\ref{tab4} into six dedicated tables~\ref{tab6}~-~\ref{tab11}:  Si containing species (Table~\ref{tab6}), SiCC (Table~\ref{tab7}), metal containing species (Table~\ref{tab8}), C and H compounds (Table~\ref{tab9}), Nitrogen containing species (Table~\ref{tab10}), O and S containing species (Table~\ref{tab11}). Unidentified lines are present in Table~\ref{tab5}. 

The columns in Tables~\ref{tab4}~-~\ref{tab11} are as follows: species
name, transition, line frequency and uncertainty 
in MHz from fitting, catalogue frequency from CDMS/JPL in MHz, integrated main beam temperature
and $1\sigma$ uncertainty in K\,km\,s$^{-1}$, main beam line center
temperature $T_{\rm mb}$ and $1\sigma$ uncertainty in mK from fitting, envelope
expansion velocity and $1\sigma$ uncertainty in km\,s$^{-1}$, and the
coding for the line blending status. If the fitted line frequency and expansion
velocity are given without uncertainty, they are set at catalogue frequency and 14.5\,km\,s$^{-1}$ respectively for
line profile fitting. This is often the case when a line is either too weak
or blended with other lines. The line frequency uncertainty given by
the truncated parabolic profile fitting method in CLASS was found to be too
small. Therefore, we merely present spectral resolution as the
uncertainty of most frequencies in
Tables~\ref{tab4}~-~\ref{tab11}. However, the frequency uncertainty of some weak lines and all
unidentified lines (also weak) were estimated by eye, and hence is usually worse than the spectral resolution. The one-letter mark that follows the catalogue frequency of some entries means the frequency is an average of several blended (hyper)fine structure components. Most of the integrated line intensities were derived by direct integration of the line spectrum, except for those entries with their integrated intensity followed by a star symbol in the tables where the integrated line intensities were derived by line profile fitting in CLASS.

Two species, $^{13}$CCH and HN$^{13}$C, are of the first detection in
\object{IRC +10216}. The isotopic hydrogen isocyanide HN$^{13}$C has
four rotational lines below 400\,GHz, namely, $J=1-0$ ($87$\,GHz),
$2-1$ ($174$\,GHz), $3-2$ ($261$\,GHz) and $4-3$ ($348$\,GHz). Due to 
the limited sensitivity of the earlier surveys, the
$J=1-0$ line was not detected by \cite{Joha85} and the $J=4-3$
line not detected by either \cite{Aver92} or \cite{Groe94}. The $J=2-1$ line was out of
the frequency coverage of the \citetalias{Cern00} survey. The $J=3-2$ line at $261.263$\,GHz is
clearly detected in our survey, with the integrated line intensity
larger than $7\sigma$. Similarly, the isotopic ethynyl $^{13}$CCH has
4 groups of fine and hyperfine structure lines below 400\,GHz, namely,
$N=1-0$ ($84$\,GHz), $2-1$ ($168$\,GHz), $3-2$ ($252$\,GHz) and $4-3$
($336$\,GHz) groups. The $N=1-0$ line group was covered by the survey
of \cite{Joha85} but not detected. The $N=2-1$ line group was within
the frequency range of \citetalias{Cern00} survey but not detected,
due to the weakness of the lines and limited S/N ratio of that portion of
data and perhaps too small beam of the IRAM 30M telescope
\citep[because $^{13}$CCH may follow its main isotopic species CCH to
show the hollow spherical distribution as given by][]{Guel96}. The
$N=4-3$ line group was covered by the survey of
\cite{Groe94}. Actually, their spectrum did show some unusual features
around the frequency $336.6$\,GHz, but the limited S/N ratio made
the identification inconclusive. The $N=3-2$ line group between
$252.4-252.6$\,GHz in our survey data shows clear detection of its
partially blended fine and hyperfine structure components.

\section{Discussions}

The line frequencies of all identified lines generally agree well with the 
predicted frequencies from CDMS or JPL. The LSR velocity of \object{IRC +10216}
calculated from the uncertainty weighted average over 49 double horn strong lines
  (better than $10\,\sigma$ detection) is $V_{\rm LSR} =
  -26.404\pm 0.004$\,km\,s$^{-1}$, close to the value of -26.5\,km\,s$^{-1}$ from
  \citetalias{Cern00} survey. The standard value of $V_{\rm LSR} = -26.5$\,km\,s$^{-1}$ was adopted for other parts of this work. The expansion velocity of the circumstellar
envelope calculated from the same sample of lines is $V_{\rm exp} =
13.61\pm 0.05$\,km\,s$^{-1}$, a little different from the value of 14.5\,km\,s$^{-1}$
from \citetalias{Cern00} survey. The difference might not be
significant, because the line profile is usually not perfectly truncated
parabola even in the optically thin case. The measurements given
here were derived by truncated parabolic profile fitting in CLASS and should be taken as FWHM of the lines. 

\subsection{A brief comparison with the survey of Cernicharo et al.}

The whole $\lambda\,2$\,mm part of our survey (131.2-160.3\,GHz)
overlaps with the survey by \citetalias{Cern00}. The main differences
of the two surveys lie in following facts: 1) Our survey were
performed in a relatively short duration (within one year) while
\citetalias{Cern00} survey were done in 11 years. 2) Our survey has
more homogeneous data quality than \citetalias{Cern00} survey because
the IRAM 30M telescope experienced major technical updates during
those 11 years. 3) Although our telescopes are smaller than
the IRAM 30M telescope, we used longer integration time, so that the
S/N is higher than \citetalias{Cern00} survey in a number of
wavelength ranges. For example, we have detected CCS, $N=11-10,
J=12-11$ around $144246$\,MHz and
{\it c}-C$_3$H$_2$,$2_{2,0}-1_{1,1}$ around $150437$\,MHz, while these lines
were buried in the high noise of the \citetalias{Cern00} survey
spectra. 4) Our surveys have larger beam sizes ($30\arcsec-40\arcsec$) than
\citetalias{Cern00} survey ($15\arcsec$), which means the two surveys
were looking at slightly different regions of the circumstellar
envelope. The large beam size in our survey has advantage in
detecting molecular species with extended distribution. For example, the inner edge of
the shell-like C$_4$H distribution region had been resolved by the
small beam of the IRAM 30M telescope in the \citetalias{Cern00}
survey, so that all the the C$_4$H emission lines became narrow double
peaks in their spectra, while in our survey, they take a look of
typical broad and strong double peak profile. Another consequence of
different beams is that the column densities of C$_3$H, C$_4$H and
C$_3$N determined from our survey data are all larger than
that from \citetalias{Cern00} survey by a factor of two (see the details in next section
and in Table~\ref{coldens}), which can be interpreted by the inclusion in our larger beams 
the bright molecular emission ring at the radius of $15\arcsec$.

Compared with the \citetalias{Cern00} survey results, lines from 18 species (or isotopomers or vibrational states) were not detected. They are CP, CS($v=1$), $^{13}$CCCN, C$^{13}$CCN, CC$^{13}$CN, C$_5$H, HC$_2$N, HC$_3$N($\nu_7=1$), HC$_5$N, KCl, K$^{37}$Cl, $^{26}$MgNC,$^{25}$MgNC, SiCC($\nu_3=1$), SiC$_3$, SiS($v=2,3$), $^{29}$Si$^{34}$S, $^{30}$Si$^{34}$S. The species SiS, CS and KCl are known to show centrally peaked emission distribution \citep{Luca95,Cern87}, hence the non-detection of $^{29}$Si$^{34}$S, $^{30}$Si$^{34}$S, SiS($v=2,3$), CS($v=1$), KCl and K$^{37}$Cl can be explained by larger beam dilution in our survey. Another three species, C$_3^{34}$S, H$_2$S and $^{29}$SiO, had been detected by \citetalias{Cern00} survey, but the frequencies were not covered by our survey. 

In the $\lambda\,1.3$\,mm survey, we detected 17 more species (or
isotopomers or vibrational states): CCH, $^{13}$CCH, C$^{13}$CH,
CN, CO, $^{13}$CO, C$^{17}$O, C$^{18}$O, H$_2$CO, $^{13}$CCCH, {\it c}-$^{13}$CCCH, H$^{13}$CN, H$^{13}$CN($\nu_2=1^1$), HC$^{15}$N, HN$^{13}$C, SiO($v=1$), compared to the \citetalias{Cern00} $\lambda\,2$\,mm survey. The main isotopic species of these molecules are all already known in \object{IRC +10216}, according to the attached molecule list in the end of Table 1 of \citetalias{Cern00}. However, according to our knowledge, it is the first time to detect the isotopic species $^{13}$CCH and HN$^{13}$C in \object{IRC +10216}, as discussed in Sect.~\ref{results}. The molecules C$_3$H$_2$ and $^{13}$CCCH$_2$ detected by both \citetalias{Cern00} and our surveys are cyclic isomers {\it c}-C$_3$H$_2$ and {\it c}-$^{13}$CCCH$_2$, according to the CDMS line list of {\it l}-C$_3$H$_2$ and {\it l}-$^{13}$CCCH$_2$ and the JPL line list of {\it c}-C$_3$H$_2$ and {\it c}-$^{13}$CCCH$_2$. 

\subsection{Column densities and rotational temperatures}
\label{columnd}

Rotational temperatures and column densities were determined for those
molecules that show extended spatial distribution in interferometry
observations. The standard formula for the rotation analysis of optically thin lines from a homogeneous medium is as follows \citep[see, e.g., the deduction of the formula in][]{Kawa95,Cumm86}:
$$\log \left( \frac{3kW} {8\pi^3\nu S\mu_{\rm i}^2} \right)=
\log \frac{N\left(T_{\rm ex}-T_{\rm bg}\right)}{QT_{\rm
    ex}}-\frac{E_{\rm up}\log e}{kT_{\rm ex}},$$
where $k$ is Boltzmann constant, $W$ is integrated line intensity, $S$ is the transition
strength, $\mu_{\rm i}$ is the permanent dipole moment along molecular
axis i (=a, b or c), $N$ is the column density, $T_{\rm ex}$ and $T_{\rm
  bg}$ are the excitation temperature and background brightness
temperature respectively, Q is the partition function, and $E_{\rm up}$ is
the energy of the upper level of the transition. The unit of $W$ is
K\,cm\,s$^{-1}$, while all other quantities are in normal cgs units. The value
of $Q$ at different temperatures and $\mu_{\rm i}$ can be found in the
documentation entries of the CDMS or JPL catalogues. The integrated line intensity $W$ has been corrected for beam dilution by using source sizes from
  literature (see in Table~\ref{coldens}) and assuming gaussian source shape. For the
  expanding shell around evolved stars, this formula gives a gross
  estimate of the column density through the whole circumstellar
  envelope (i.e., two times of the radial column density).
Almost all molecules involved in this work were present in either CDMS or JPL or both, except metal cyanides and SiC$_2$ isotopologues that were computed during this work.  

The derived excitation temperatures ($T_{\rm ex}$) and column densities ($N_{\rm col}$),
 together with some results from the literature, are summarized in
  Table~\ref{coldens}. The agreement of the column densities with
  literature results is generally good. Larger discrepancy is found for {\it c}-C$_3$H$_2$ and SiCC. There are several major sources of uncertainty in $T_{\rm ex}$ and $N_{\rm col}$. Although in the rotation diagram analysis we have excluded
  several lines that are obviously weaker than expected from
  optically thin case (low transitions) or from LTE condition (high transitions), weaker opacity effect or
  non-LTE effect are still very difficult to
  identify and remove. The non-gaussian source brightness distribution (e.g., ring-like pattern) will result in a beam dilution behavior different from gaussian source if observations are done with different beam sizes. Furthermore, the source size variation
  from transition to transition has been neglected. Another source of
  uncertainty is related to the 
  fact that our beam size is comparable to the source sizes. Our
  beam size in 1.3\,mm survey is about $30\arcsec$ (comparable to the size of most
  of the ring-like distributions) while our 2\,mm survey has larger beam size of
  $40\arcsec$. This difference means that the 2\,mm survey may has covered
  most of the ring-like emission region while the 1.3\,mm survey may
  has missed a significant part of it, which may introduce additional
  uncertainty into $T_{\rm ex}$ and $N_{\rm col}$ in
  some cases. 

The derived ortho/para ratio of {\it c}-C$_3$H$_2$ is roughly 4, close to the expected theoretical value of 3. The column densities of C$_4$H in the vibrationally excited states $\nu_7=1$ and $2^0$ are only slightly lower than in the ground state, which indicates strong over-population in the vibrational states because all the vibrational levels have energy (about 300\,K) well above the typical kinematic temperature (around 50\,K) in the circumstellar envelope of \object{IRC +10216}. This could be explained by infrared radiation excitation of the vibrational levels. 

\begin{figure}
\epsscale{0.50}
\plotone{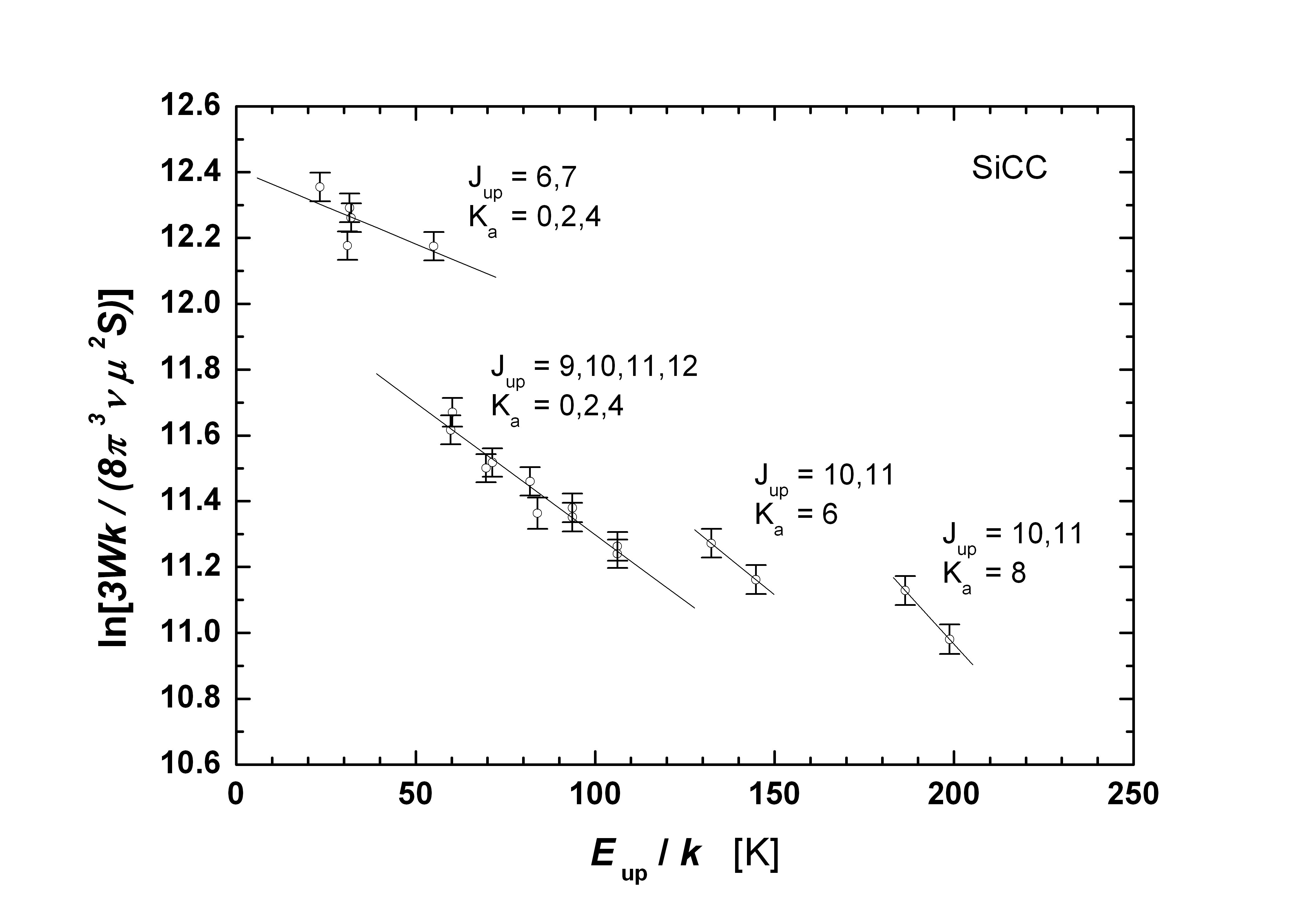}
\caption{Rotation diagram of SiCC lines. A relative calibration uncertainty of $10\%$ has been included in the error bar. \label{figsicc}}
\end{figure}

The rotation diagram of the slightly asymmetric molecule SiCC is
unusual. The different transitions with lower and higher $J_{\rm
  up}$ or with different quantum number $K_a$ are located along
different more or less parallel straight line segments (See Fig.~\ref{figsicc}). The unusual rotation diagram pattern of this molecule was first addressed by \cite{Thad84}, and later elaborated by \cite{Aver92} with more high frequency observations. The latter authors argued that there might be three different temperature components in the rotation diagram of SiCC: $T_{\rm ex} = 14$\,K in lower $K_a$ ladders, 60\,K in higher $K_a$ ladders, and 145-227\,K across $K_a$ ladders. The reason for the quite different intra-$K_a$-ladder and inter-$K_a$-ladder rotational temperatures is that the radiation transition probability between the $K_a$ ladders is much weaker than within each $K_a$ ladders, so that the relative population between the $K_a$ ladders might be controlled by collision, while that within each $K_a$-ladder might be controlled by spontaneous radiation decay. However, a kinematic temperature of $>140$\,K seems to be too high for the gas at the large radius where SiCC appears. So the infrared excitation mechanism suggested by \cite{Aver92} might be a better interpretation of the high excitation temperature across $K_a$ ladders than collisional excitation. This is supported by the tentative detection rotational of transitions in the vibrationally excited $\nu_3=1$ antisymmetric mode of SiCC in \object{IRC +10216} by \cite{Gens97} using the transition frequency list from \cite{Boge91}. The different rotation temperatures in the lower and higher $K_a$ ladders could be interpreted by two components in the density distribution: a ring-like extended component and a centrally peaked compact component \citep[see the map in ][]{Luca95}. There are some other papers \citep[e.g.,][]{Groe94,Kawa95} that gave quite different rotation temperatures and column densities (see in Table~\ref{coldens}). Our results are also different from the earlier results. Particularly, we have some points in the transitional region between the low and high energy levels (around $E/k = 70$\,K in Fig.~\ref{figsicc}). These points (with $K_a=0, 2, 4$ and $J_{\rm up}=9,10,11,12$) roughly align along the same straight line, which indicates that the inter-$K_a$ ladder temperatures may not be so different from the intra-$K_a$ ladder temperatures among these energy levels. More detailed analysis is needed to interpret this phenomenon.

\subsection{Isotopic ratios}

Elemental isotopic ratio can be determined from intensity ratio of optically thin lines of isotopic species pairs \citep[see discussions by][]{Kaha88,Kaha92}. But the poorly known effects of opacity, beam dilution, excitation and isotopic fractionation on the line intensity ratios make the accurate determination of isotopic ratio difficult. Here, for most lines, we simply neglect these effects, and take the integrated line intensity ratio of possibly optically thin lines, after correction for the frequency dependence of $\nu^{2}$ (for both line intensity in unit of K and integrated line intensity in unit of K\,km\,s$^{-1}$), as the isotopic ratio. However, for the case of $^{13}$CO, $J=2-1$ line, opacity effect has been checked by channel-by-channel profile comparison (see details below). The relative calibration uncertainty of $10\%$ has been taken into account in these calculations.

\subsubsection{Oxygen and carbon isotopes}

\begin{figure}
\epsscale{0.50}
\plotone{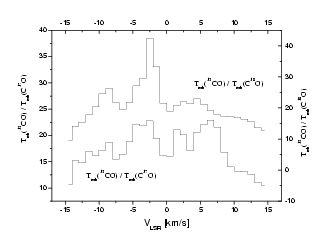}
\caption{The channel-by-channel $J=2-1$ line intensity ratios $T_{\rm mb}$($^{13}$CO)/$T_{\rm mb}$(C$^{17}$O) and $T_{\rm mb}$($^{13}$CO)/$T_{\rm mb}$(C$^{18}$O) against LSR velocity $(-14.5,+14.5)$\,km\,s$^{-1}$ that show opacity effects on both edges of the profile (the drop-down trends).  \label{profratio}}
\end{figure}

Table~\ref{isotope} shows that our $^{12}$C/$^{13}$C ratio ($34.7\pm 4.3$) determined from [SiCC]/[Si$^{13}$CC] is slightly smaller than the best value of $47(+6,-5)$ from \cite{Kaha88}. We have noticed that the SiCC lines are strong and hence may suffer from opacity effect. But the Si$^{13}$CC lines are too noisy to allow channel-by-channel comparison. Therefore, our smaller $^{12}$C/$^{13}$C ratio could be attributed to opacity effect. The opacity effect in $^{13}$CO line profile is also obvious. A channel-by-channel comparison introduced by \cite{Kaha92} is performed between the $J=2-1$ line profile of $^{13}$CO and that of C$^{17}$O and C$^{18}$O (see Fig.~\ref{profratio}). The drop-down trend near both edges of the line profiles demonstrates the opacity effect near the line edges of $^{13}$CO $J=2-1$ line. Therefore, only the central portion between (-5,+5)\,km\,s$^{-1}$ is averaged to compute the line intensity ratios. In the calculation of $^{16}$O/$^{17}$O and $^{16}$O/$^{18}$O ratios from double ratios [$^{13}$CO]/[C$^{17}$O] and [$^{13}$CO]/[C$^{18}$O], we do not use our $^{12}$C/$^{13}$C value, but use $47\pm 5.5$ from \cite{Kaha88}. Eventually, with the opacity effect removed, our $^{16}$O/$^{17}$O and $^{16}$O/$^{18}$O ratios agree quite well with that of \cite{Kaha92}. Therefore, our results confirm the conclusion that the isotopic ratios of S and Si are nearly solar while that of C and O are non-solar in \object{CW Leo} (see the comparison with solar values in Table~\ref{isotope}).

\subsubsection{Sulfur and silicon isotopes}

The abundance ratios of the rare isotopes of sulfur and silicon, $^{33}$S, $^{34}$S, $^{29}$Si and $^{30}$Si, are determined from CS, SiCC, SiS and some double ratios of SiS (see in the last column of Table~\ref{isotope}). The derived results are similar to the results from \cite{Kaha88} and so all isotopic ratios of S and Si are confirmed to be close to the solar values. 

\subsubsection{Abundance differences among $^{13}$C isotopomers}

We investigated possible abundance differences among $^{13}$C
  substituted species of HC$_3$N. The integrated line intensity ratios
  (with the factor $\nu^2$ corrected and 10\% of relative calibration
  error included) H$^{13}$CCCN:HC$^{13}$CCN:HCC$^{13}$CN
  $=1.00:1.19(\pm 0.14):1.31(\pm 0.15)$ were determined from two
  transitions $J=16-15$ and $17-16$, with very similar values from the
  two lines. Another transition $J=15-14$ was not used because this
  line of H$^{13}$CCCN was observed in poor weather (with atmospheric
  opacity $\tau=0.65$). Our observations suggest that the abundances
  of the $^{13}$C substituted species of HC$_3$N are not equal to each
  other in \object{IRC +10216}, contradicting with the results of
  \cite{Kaha88} (their ratios can be reproduced from their given
  intensities as $1.00 : 0.94(\pm 0.16) : 0.94(\pm 0.18)$). The
  possible reason might be the smaller beam of IRAM 30M telescope used
  by Kahane et al. The abundance differentiation among the three $^{13}$C
  variants of HC$_3$N is argued by \cite{Taka98} to be enhanced in the low-temperature
regions.   Our observations with a larger beam probed the cooler
outer parts of the CSE. Moreover, our result of the HC$_3$N isotopic ratios in the carbon star is in the same trend as the corresponding ratios of $1.0:1.0:1.4$ found in interstellar cloud \object{TMC-1} by \cite{Taka98}, i.e., the $^{13}$C atom is more concentrated in the C atom adjacent to the N atom. 

An integrated line intensity ratio $^{13}$CCH/C$^{13}$CH$=1.00:0.81(\pm0.09)$ was determined from their
 $N=3-2$ lines. However, the slightly higher abundance of $^{13}$CCH
 may not be true, because not only the $^{13}$CCH line in our survey
 was observed in a relatively worse weather (with atmospheric opacity
 $\tau=0.5$), but also some of the weak hyperfine structure components may
 have been left out when integrating the complex blended hyperfine
 structure line profiles. \cite{Taka98} mentioned the possible
 abundance difference between $^{13}$CCH and C$^{13}$CH in the
 \object{Orion A} ridge that was found by \cite{Sale94} (the original
 paper unavailable). However, the calibration was said to be a problem
 in that work. Therefore, the abundance differenciation between
 $^{13}$CCH and C$^{13}$CH is still inconclusive in either molecular clouds or circumstellar envelope.

\subsection{Other interesting features}

We have detected rotational lines from vibrationally excited states of five
molecules: C$_3$H, C$_4$H, HCN, H$^{13}$CN and SiS. C$_4$H shows
strong rotational lines from the $\nu_7=1$, $2^0$
and $2^2$ states. The large column densities derived from rotation diagram
analysis of these states in Sect.~\ref{columnd} indicate that infrared
pumping could be very efficient. We have detected rotational lines from
the excited bending states 
$\nu_2 = 1^1,2^0,3^1$ of HCN. The two well separated $\ell$-doublets
of H$^{13}$CN in the excited $\nu_2=1^1$ bending mode both show clear
triangular profiles, which indicates that the line emission comes from
the inner acceleration region of the circumstellar envelope. The two
lines from the $\nu_4=1$ state of C$_3$H are weak and their
detections are tentative. The line from the 
$v=1$ state of SiS is also very weak, perhaps due to large beam
dilution.

Four very narrow lines have been found in the survey. One of them
  is the $\nu_{2}=3^{1e},J=3-2$ line of HCN, while the other
  three are from the $v=0,1$ states of SiS (see the list in
  Table~\ref{narrowlines}). There are two possible interpretations for these narrow line widths: maser
  beaming effect or thermal emission from the wind acceleration zone where
  outflow velocity is not fully developed. For example, the peak mainbeam
  temperature of the narrow HCN line in our SMT beam (FWHM $= 28\arcsec$) with a resolution
  of 1\,MHz was 96\,mK. If we assume it arises within 10 stellar
  radii ($\approx 4\times 10^{14}$\,cm, assuming a luminosity of
  $10^4$\,L$_{\sun}$ and surface temperature of 2500\,K for the star),
  with the distance of 120\,pc to the star, the beam-dilution corrected
  brightness temperature is about 380\,K, reasonable for a line thermally excited in a hot region near the
  photosphere. On the other hand, if the line arises
  from a local region smaller than the stellar size, then the
  brightness temperature would be $>3.8\times 10^4$\,K,
  consistent with a weak maser that is magnifying the stellar continuum
  emission. In both cases, these lines could be varying in intensity with the
  pulsation of the central star. The observation
  dates and other parameters (observed frequency, mainbeam temperature
  at the peak and FWHM line width) are given in Table~\ref{narrowlines} for a follow-up
  study. The HCN line is the first detection from its $\nu_2=3^{1}$ vibrational state in \object{CW Leo}. The SiS, $v=0,J=14-13$ line had been
  observed by \cite{Fonf06} and was argued to be masers
  superposed on the top of the broad thermal line.  \cite{Turn87}
  reported the detection of both SiS $v=1$ lines (with lower S/N ratio) and
  argued that the $J=13-12$ line may be a thermal line while the sharp
  peak of $J=14-13$ line could be a weak maser. The blue $J=14-13$
  maser peak in our spectrum 
  shows a velocity shift of about -7.4\,km\,s$^{-1}$ with respect to the stellar
  velocity of -26.5\,km\,s$^{-1}$. The pumping mechanism of the SiS masers could be mid-infrared line-overlapping with C$_2$H$_2$, HCN, and their $^{13}$C isotopologues, as suggested by \cite{Fonf06}.

Clear triangular line profiles were found for three lines: CCS, $N=15-14, J=15-14$ (234814.63\,MHz) and the two well separated $\ell$-doublets in the excited bending state of H$^{13}$CN, $\nu_2=1^1,J=3-2$ (258936.82 and 260225.42\,MHz). These triangular lines might arise from the accelerating inner shell of the circumstellar envelope where the terminal outflow velocity is not fully developed yet.

\section{Summary}

We have surveyed the circumstellar envelope of the archetypal carbon
rich AGB star \object{IRC +10216} in $\lambda\,1.3$ and 2\,mm ranges
using the KP12M and SMT telescopes of ARO. In total,
  377 lines were detected and 26 molecules plus 29 rare isotopic
  species and 2 cyclic isomers were identified to be carriers for 360
  of them. Only 17 weak lines remain unidentified. 

The rotational lines of two isotopic species $^{13}$CCH and HN$^{13}$C
were detected in \object{IRC +01216} for the first time by our survey. Two rotation lines of the formaldehyde have been
confirmed. Column densities of 15 extended species roughly agree with
previous determinations. Isotopic ratios of C, O, S and Si confirm the
conclusion that the S and Si isotopic ratios are nearly solar while
the C and O isotopic ratios are non-solar. Our data suggests possible
abundance differentiation among the three $^{13}$C substituted
isotopic isomers of HC$_3$N.

Modified spectroscopic parameters of alkali metal cyanide NaCN/NaNC, Na$^{13}$CN/NaN$^{13}$C, KCN/KNC and silacyclopropynylidene SiC$_2$ have been derived to help line identification. Transition frequencies from the current surveys were used to improve the spectroscopic parameters of Si$^{13}$CC, $^{29}$SiC$_2$ and $^{30}$SiC$_2$.

\acknowledgments

We thank the referee, Dr. M. Gu{\'e}lin, for many pertinent comments
that have greatly improved this work. J.H. thanks Dr. H.M. Pickett for the help on understanding the JPL
catalogue, Dr. William Peters for the help on the calibration of the
spectral data, Dr. Mao Rui-Qing for discussions on instrumental
effects in our spectra, Mr. Nico Koning for discussions on line identification, and the GILDAS help desk for the help on using CLASS. We also thank ARO telescope operators for the assistance in remote observations. H.S.P.M. and the CDMS are supported by the Bundesministerium f\"ur Bildung und Forschung (BMBF)
 administered through Deutsches Zentrum f\"ur Luft- und Raumfahrt (DLR; the German space agency). J.H. also thanks the projects No. 10433030 and 10503011 of the National Natural Science Foundation of China. T.H. acknowledges the support from NSC grant NSC 96-2112-M-001-018-MY3.

{\it Facilities:} \facility{HHT ()}, \facility{NRAO:12m ()}.

\bibliographystyle{aa}  
\bibliography{ms,lab_bib}   

\begin{thebibliography}{7}

\bibitem[{{van Vaals} {et~al.}(1984){van Vaals}, {Meerts}, \& {Dymanus}}]{NaCNrot}
{van Vaals}, J.~J., {Meerts}, W.~L., \& {Dymanus}, A.
1984, J. Chem. Phys., 86, 147

\bibitem[{{Kuijpers} {et~al.}(1976){Kuijpers}, {T\"orring}, \& {Dymanus}}]{KCNrot1}
{Kuijpers}, P., {T\"orring}, T. \& {Dymanus} A.
1976, Chem. Phys. Lett., 42, 423

\bibitem[{{T\"orring} {et~al.}(1980){T\"orring}, {Bekooy}, {Meerts}, {Hoeft} 
{Tiemann} \& {Dymanus}}]{KCNrot2}
{T\"orring}, T., {Bekooy}, J.~P., {Meerts}, L.~W., {Hoeft} J., {Tiemann}, E. 
\& {Dymanus}, A.,
1980, J. Chem. Phys., 73, 4875

\bibitem[{{van Vaals} {et~al.}(1984){van Vaals}, {Meerts}, \& {Dymanus}}]{KCNrot3}
{van Vaals}, J.~J., {Meerts}, W.~L., \& {Dymanus}, A.
1984, J. Mol. Spectrosc., 106, 280

\bibitem[{Suenram et al.(1989){Suenram}, {Lovas}, \& {Matsumura}}]{SiC2_1-0}
Suenram, R.~D., Lovas, F.~J., \& Matsumura, K. 
1989, ApJ, 342, L103 

\bibitem[Gottlieb et al.(1989)]{SiC2_rot} 
Gottlieb, C.~A., Vrtilek, J.~M., \& Thaddeus, P. 
1989, ApJ, 343, L29 

\bibitem[{Cernicharo et al.(1991){Cernicharo}, {Guelin}, {Kahane}, {Bogey}, \& {Demuynck}}]{SiC2_isos} 
Cernicharo, J., Guelin, M., Kahane, C., Bogey, M., \& Demuynck, C. 
1991, A\&A, 246, 213 

\end{thebibliography}


\begin{thebibliography}{39}
\expandafter\ifx\csname natexlab\endcsname\relax\def\natexlab#1{#1}\fi

\bibitem[{{Avery} {et~al.}(1992){Avery}, {Amano}, {Bell}, {Feldman}, {Johns},
  {MacLeod}, {Matthews}, {Morton}, {Watson}, {Turner}, {Hayashi}, {Watt}, \&
  {Webster}}]{Aver92}
{Avery}, L.~W., {Amano}, T., {Bell}, M.~B., {et~al.} 1992, \apjs, 83, 363

\bibitem[{{Bieging} \& {Tafalla}(1993)}]{Bieg93}
{Bieging}, J.~H. \& {Tafalla}, M. 
1993, \aj, 105, 576

\bibitem[{{Bogey} {et~al.}(1991){Bogey}, {Demuynck}, {Destombes}, \&
  {Walters}}]{Boge91}
{Bogey}, M., {Demuynck}, C., {Destombes}, J.~L., \& {Walters}, A.~D. 
1991, \aap, 247, L13

\bibitem[{{Brown} {et~al.}(1975){Brown}, {Hougen}, {Huber}, {Johns}, {Kopp},
  {Lefebvre-Brion}, {Merer}, {Ramsay}, {Rostas}, \& {Zare}}]{Brow75}
{Brown}, J.~M., {Hougen}, J.~T., {Huber}, K.-P., {et~al.} 
1975, J. Mol. Spectrosc., 55, 500

\bibitem[{{Cernicharo} \& {Gu{\'e}lin}(1987)}]{Cern87}
{Cernicharo}, J. \& {Gu{\'e}lin}, M. 1987, \aap, 183, L10

\bibitem[{Cernicharo et al.(1991){Cernicharo}, {Gu{\'e}lin}, {Kahane}, {Bogey}, \& {Demuynck}}]{SiC2-isos} 
Cernicharo, J., Gu{\'e}lin, M., Kahane, C., Bogey, M., \& Demuynck, C. 
1991, A\&A, 246, 213 

\bibitem[{{Cernicharo} {et~al.}(2000){Cernicharo}, {Gu{\'e}lin}, \&
  {Kahane}}]{Cern00}
{Cernicharo}, J., {Gu{\'e}lin}, M., \& {Kahane}, C. 
2000, \aaps, 142, 181

\bibitem[{{Cherchneff} {et~al.}(1993){Cherchneff}, {Glassgold}, \&
  {Mamon}}]{Cher93}
{Cherchneff}, I., {Glassgold}, A.~E., \& {Mamon}, G.~A. 
1993, \apj, 410, 188

\bibitem[{{Cummins} {et~al.}(1986){Cummins}, {Linke}, \& {Thaddeus}}]{Cumm86}
{Cummins}, S.~E., {Linke}, R.~A., \& {Thaddeus}, P. 
1986, \apjs, 60, 819

\bibitem[{{Fonfr{\'{\i}}a Exp{\'o}sito} {et~al.}(2006){Fonfr{\'{\i}}a
  Exp{\'o}sito}, {Ag{\'u}ndez}, {Tercero}, {Pardo}, \& {Cernicharo}}]{Fonf06}
{Fonfr{\'{\i}}a Exp{\'o}sito}, J.~P., {Ag{\'u}ndez}, M., {Tercero}, B.,
  {Pardo}, J.~R., \& {Cernicharo}, J. 
2006, \apjl, 646, L127

\bibitem[{{Ford} {et~al.}(2004){Ford}, {Neufeld}, {Schilke}, \&
  {Melnick}}]{Ford04}
{Ford}, K.~E.~S., {Neufeld}, D.~A., {Schilke}, P., \& {Melnick}, G.~J. 2004,
  \apj, 614, 990

\bibitem[{{Gensheimer} \& {Snyder}(1997)}]{Gens97}
{Gensheimer}, P.~D. \& {Snyder}, L.~E. 
1997, \apj, 490, 819

\bibitem[Gottlieb et al.(1989)]{SiC2-rot} 
Gottlieb, C.~A., Vrt{\'{\i}}lek, J.~M., \& Thaddeus, P. 
1989, ApJ, 343, L29 

\bibitem[{{Groenewegen} {et~al.}(1998){Groenewegen}, {van der Veen}, \&
  {Matthews}}]{Groe98}
{Groenewegen}, M.~A.~T., {van der Veen}, W.~E.~C.~J., \& {Matthews}, H.~E.
1998, \aap, 338, 491

\bibitem[{{Groesbeck} {et~al.}(1994){Groesbeck}, {Phillips}, \&
  {Blake}}]{Groe94}
{Groesbeck}, T.~D., {Phillips}, T.~G., \& {Blake}, G.~A. 
1994, \apjs, 94, 147

\bibitem[{{Gu{\'e}lin} {et~al.}(1993){Gu{\'e}lin}, {Lucas}, \& {Cernicharo}}]{Guel93}
{Gu{\'e}lin}, M., {Lucas}, R., \& {Cernicharo}, J. 
1993, \aap, 280, L19

\bibitem[{{Gu{\'e}lin} {et~al.}(1996){Gu{\'e}lin}, {Lucas}, \& {Neri}}]{Guel96}
{Gu{\'e}lin}, M., {Lucas}, R., \& {Neri}, R. 1996, in IAU Symposium, Vol. 170, CO:
  Twenty-Five Years of Millimeter-Wave Spectroscopy, ed. W.~B. {Latter}, J.~E.
  {Radford Simon}, P.~R. {Jewell}, J.~G. {Mangum}, \& J.~{Bally}, 359

\bibitem[{{Johansson} {et~al.}(1985){Johansson}, {Andersson}, {Elder},
  {Friberg}, {Hjalmarson}, {Hoglund}, {Olofsson}, {Rydbeck}, \&
  {Irvine}}]{Joha85}
{Johansson}, L.~E.~B., {Andersson}, C., {Elder}, J., {et~al.} 
1985, \aaps, 60, 135

\bibitem[{{Kahane} {et~al.}(1992){Kahane}, {Cernicharo}, {Gomez-Gonzalez}, \&
  {Gu{\'e}lin}}]{Kaha92}
{Kahane}, C., {Cernicharo}, J., {Gomez-Gonzalez}, J., \& {Gu{\'e}lin}, M. 
1992, \aap, 256, 235

\bibitem[{{Kahane} {et~al.}(1988){Kahane}, {Gomez-Gonzalez}, {Cernicharo}, \&
  {Gu{\'e}lin}}]{Kaha88}
{Kahane}, C., {Gomez-Gonzalez}, J., {Cernicharo}, J., \& {Gu{\'e}lin}, M. 
1988, \aap, 190, 167

\bibitem[{{Kawaguchi} {et~al.}(1995){Kawaguchi}, {Kasai}, {Ishikawa}, \&
  {Kaifu}}]{Kawa95}
{Kawaguchi}, K., {Kasai}, Y., {Ishikawa}, S.-I., \& {Kaifu}, N. 
1995, \pasj, 47, 853

\bibitem[{{Kuijpers} {et~al.}(1976){Kuijpers}, {T\"orring}, \& {Dymanus}}]{KCNrot1}
{Kuijpers}, P., {T\"orring}, T. \& {Dymanus} A.
1976, Chem. Phys. Lett., 42, 423

\bibitem[{{Lovas}(2004)}]{Lova04}
{Lovas}, F.~J. 
2004, J. Phys. Chem. Ref. Data, 33, 177

\bibitem[{{Lucas} \& {Gu{\'e}lin}(1999)}]{Luca99}
{Lucas}, R. \& {Gu{\'e}lin}, M. 1999, in IAU Symposium, Vol. 191, Asymptotic
  Giant Branch Stars, ed. T.~{Le Bertre}, A.~{Lebre}, \& C.~{Waelkens}, 305

\bibitem[{{Lucas} {et~al.}(1995){Lucas}, {Gu{\'e}lin}, {Kahane}, {Audinos}, \&
  {Cernicharo}}]{Luca95}
{Lucas}, R., {Gu{\'e}lin}, M., {Kahane}, C., {Audinos}, P., \& {Cernicharo}, J.
1995, \apss, 224, 293

\bibitem[{{Millar} \& {Herbst}(1994)}]{Mill94}
{Millar}, T.~J. \& {Herbst}, E. 
1994, \aap, 288, 561

\bibitem[{{Millar} {et~al.}(2000){Millar}, {Herbst}, \& {Bettens}}]{Mill00}
{Millar}, T.~J., {Herbst}, E., \& {Bettens}, R.~P.~A. 
2000, \mnras, 316, 195

\bibitem[{{M{\"u}ller} {et~al.}(2005){M{\"u}ller}, {Schl{\"o}der}, {Stutzki},
  \& {Winnewisser}}]{Muel05}
{M{\"u}ller}, H.~S.~P., {Schl{\"o}der}, F., {Stutzki}, J., \& {Winnewisser}, G.
2005, J. Mol. Struct., 742, 215

\bibitem[{{M{\"u}ller} {et~al.}(2001){M{\"u}ller}, {Thorwirth}, {Roth}, \&
  {Winnewisser}}]{Muel01}
{M{\"u}ller}, H.~S.~P., {Thorwirth}, S., {Roth}, D.~A., \& {Winnewisser}, G.
2001, \aap, 370, L49

\bibitem[{{Pickett} {et~al.}(1998){Pickett}, {Poynter}, {Cohen}, {Delitsky},
  {Pearson}, \& {M{\"u}ller}}]{Pick98}
{Pickett}, H.~M., {Poynter}, R.~L., {Cohen}, E.~A., {et~al.} 
1998, J. Quant.  Spectrosc. Radiat. Transfer, 60, 883

\bibitem[{{Saleck} {et~al.}(1994){Saleck}, {Simon}, {Winnewisser}, \&
  {Wouterloot}}]{Sale94}
{Saleck}, A.~H., {Simon}, R., {Winnewisser}, G., \& {Wouterloot}, J.~G.~A.
1994, Can. J.  Phys., 72, 747

\bibitem[{Suenram et al.(1989){Suenram}, {Lovas}, \& {Matsumura}}]{SiC2-1-0}
Suenram, R.~D., Lovas, F.~J., \& Matsumura, K. 
1989, ApJ, 342, L103 

\bibitem[{{Takano} {et~al.}(1998){Takano}, {Masuda}, {Hirahara}, {Suzuki},
  {Ohishi}, {Ishikawa}, {Kaifu}, {Kasai}, {Kawaguchi}, \& {Wilson}}]{Taka98}
{Takano}, S., {Masuda}, A., {Hirahara}, Y., {et~al.} 1998, \aap, 329, 1156

\bibitem[{{Thaddeus} {et~al.}(1984){Thaddeus}, {Cummins}, \& {Linke}}]{Thad84}
{Thaddeus}, P., {Cummins}, S.~E., \& {Linke}, R.~A. 
1984, \apjl, 283, L45

\bibitem[{{T\"orring} {et~al.}(1980){T\"orring}, {Bekooy}, {Meerts}, {Hoeft} 
{Tiemann} \& {Dymanus}}]{KCNrot2}
{T\"orring}, T., {Bekooy}, J.~P., {Meerts}, L.~W., {Hoeft} J., {Tiemann}, E. 
\& {Dymanus}, A.,
1980, J. Chem. Phys., 73, 4875

\bibitem[{{Townes} \& {Schawlow}(1955)}]{Town55}
{Townes}, C.~H. \& {Schawlow}, A.~L. 1955, {Microwave Spectroscopy} (Microwave
  Spectroscopy, New York: McGraw-Hill, 1955)

\bibitem[{{van Vaals} {et~al.}(1984a){van Vaals}, {Meerts}, \& {Dymanus}}]{NaCNrot}
{van Vaals}, J.~J., {Meerts}, W.~L., \& {Dymanus}, A.
1984, J. Chem. Phys., 86, 147

\bibitem[{{van Vaals} {et~al.}(1984b){van Vaals}, {Meerts}, \& {Dymanus}}]{KCNrot3}
{van Vaals}, J.~J., {Meerts}, W.~L., \& {Dymanus}, A.
1984, J. Mol. Spectrosc., 106, 280

\bibitem[{{Turner}(1987)}]{Turn87}
{Turner}, B.~E. 1987, \aap, 183, L23

\end{thebibliography}

\clearpage
\begin{table}[tbh]
\scriptsize 
\centering
\caption{Spectroscopic parameters\tablenotemark{a} (MHz) of alkali metal cyanides}
\label{MCN-parameters}
\begin{tabular}{lr@{}lr@{}lr@{}l} 
\hline\hline 
Parameter & \multicolumn{2}{c}{NaCN} & \multicolumn{2}{c}{Na$^{13}$CN} & 
 \multicolumn{2}{c}{KCN} \\
\hline
$A-(B+C)/2$          & 50101&.74757\,(166)   & 48021&.9343\,(33)    & 53528&.08107\,(268)  \\
$(B+C)/2$            &  7820&.190252\,(259)  &  7652&.41519\,(39)   &  4737&.751389\,(93)  \\
$(B-C)/4$            &   274&.1525055\,(177) &   272&.798660\,(39)  &   100&.959727\,(64)  \\
$D_K \times 10^3$    &   196&.52\,(169)      &   271&.79\,(169)     &   956&.79\,(247)     \\
$D_{JK} \times 10^3$ &   800&.23\,(34)       &   738&.85\,(41)      &   374&.609\,(50)     \\
$D_J \times 10^3$    &    13&.3361\,(160)    &    13&.2404\,(186)   &     5&.24462\,(209)  \\
$d_1 \times 10^3$    &  $-$2&.34908\,(34)    &  $-$2&.42697\,(110)  &  $-$0&.51452\,(78)   \\
$d_2 \times 10^3$    &  $-$1&.408951\,(182)  &  $-$1&.380792\,(296) &  $-$0&.237559\,(199) \\
$H_K \times 10^3$    &     1&.428\,(220)     &     1&.354\,(101)    &      &$-$            \\
$H_{KJ} \times 10^6$ &   155&.9(184)         &   140&.3\,(73)       &   101&.23\,(107)     \\
$H_{JK} \times 10^6$ &  $-$3&.54(120)        &  $-$2&.80\tablenotemark{b}        &  $-$7&.277\,(111)    \\
$H_J \times 10^6$    &     0&.541\,(50)      &     0&.50\tablenotemark{b}        &      &$-$            \\
$h_2 \times 10^9$    &      &$-$             &      &$-$            &  $-$5&.46\,(74)      \\
$h_3 \times 10^9$    &    39&.9\,(69)        &    30&.\tablenotemark{b}          &      &$-$            \\
\hline 
\end{tabular}
\tablenotetext{a}{Numbers in parentheses are one standard deviation in units of the 
least significant figures. Watson's $S$-reduction was used in the representation 
$I^r$.} 
\tablenotetext{b}{Estimated and kept fixed in the fits, see section~\ref{lineiden}.}
\end{table}
\begin{table}[tbH]
\scriptsize 
\centering
\caption{Spectroscopic parameters\tablenotemark{a} (MHz) of SiC$_2$ isotopic species}
\label{SiC2-parameters}
\begin{tabular}{lr@{}lr@{}lr@{}lr@{}l} 
\hline\hline 
Parameter & \multicolumn{2}{c}{SiC$_2$} & \multicolumn{2}{c}{Si$^{13}$CC} & 
 \multicolumn{2}{c}{$^{29}$SiC$_2$} & \multicolumn{2}{c}{$^{30}$SiC$_2$}  \\
\hline
$A-(B+C)/2$               &  40674&.191\,(100)        &  38928&.92\,(73)          &  40845&.6\tablenotemark{b}   &  41005&.5\tablenotemark{b}   \\
$(B+C)/2$                 &  11800&.11861\,(99)       &  11527&.5128\,(123)       &  11628&.7544\,(40)           &  11468&.7891\,(40)           \\
$(B-C)/4$                 &    678&.41301\,(130)      &    672&.5424\,(76)        &    659&.1312\,(127)          &    641&.4564\,(212)          \\
$D_K$                     &   $-$1&.2880\,(227)       &   $-$1&.223\,(65)         &   $-$1&.288\tablenotemark{b} &   $-$1&.288\tablenotemark{b} \\
$D_{JK}$                  &      1&.624219\,(88)      &      1&.540958\,(232)     &      1&.57360\,(209)         &      1&.53715\,(170)         \\
$D_J \times 10^3$         &   $-$1&.1845\,(115)       &   $-$1&.462\,(55)         &   $-$0&.554\,(139)           &   $-$0&.859\,(203)           \\
$d_1 \times 10^3$         &   $-$2&.4392\,(111)       &   $-$2&.4413\,(116)       &   $-$2&.34\tablenotemark{b}  &   $-$2&.24\tablenotemark{b}  \\
\hline 
\end{tabular}
\tablenotetext{a}{Numbers in parentheses are one standard deviation in units of the 
least significant figures. Watson's $S$-reduction was used in the representation 
$I^r$.} 
\tablenotetext{b}{Estimated and kept fixed in the fits; see section~\ref{lineiden}.}
\end{table}
\clearpage

\begin{deluxetable}{llccllcc}
\tabletypesize{\scriptsize}
\tablecaption{Summary of molecular species detected in the $\lambda\,1.3$ and 2\,mm spectral survey towards IRC +10216. \label{moltab}}
\tablewidth{0pt}
\tablehead{
\colhead{Mol} & 
\colhead{Variants} & 
\colhead{Lines} & 
\colhead{Table} &
\colhead{Mol} & 
\colhead{Variants} & 
\colhead{Lines} & 
\colhead{Table} 
}
\startdata
AlCl        & AlCl               &  3  & 8                   &HCN        & HCN                    &  1  &10 \\ 
AlCl        & Al$^{37}$Cl        &  4  & 8                   &HCN        & HCN,$\nu_2=1^1$        &  2  &10 \\  
AlF         & AlF                &  3  & 8                   &HCN        & HCN,$\nu_2=2^0$        &  1  &10 \\  
C$_2$H      & C$_2$H             &  5  & 9                   &HCN        & HCN,$\nu_2=3^1$        &  1  &10 \\  
C$_2$H      & $^{13}$CCH         &  5  & 9                   &HCN        & H$^{13}$CN             &  1  &10 \\ 
C$_2$H      & C$^{13}$CH         &  7  & 9                   &HCN        & H$^{13}$CN,$\nu_2=1^1$ &  2  &10 \\ 
C$_2$S      & C$_2$S             &  7  &11                   &HCN        & HC$^{15}$N             &  1  &10 \\ 
CH$_3$CN    & CH$_3$CN           & 14  &10                   &HC$_3$N    & HCCCN                  &  7  &10 \\
\enddata
\tablenotetext{*}{The hyperfine structure components of different fine structure 
lines of CN are partially blended. Here, only fine structure 
lines are counted as lines.}
\end{deluxetable}

\clearpage
\begin{deluxetable}{lllllrrrrrrr}
\tabletypesize{\scriptsize}
\tabletypesize{\tiny}
\tablecaption{\small Line parameters of all detected lines in our survey. The units are MHz for $\nu_{\rm obs}$ and $\nu_{\rm cat}$, K\,km\,s$^{-1}$ for $I_{\rm int}$, mK for $T_{\rm mb}$, and km\,s$^{-1}$ for $V_{\rm exp}$; the same for corresponding uncertainties. \label{tab4}}
\tablewidth{0pt}
\tablehead{
\colhead{Species} & 
\colhead{Trans} & 
\colhead{$\nu_{\rm obs}$} & 
\colhead{$\sigma$} &
\colhead{$\nu_{\rm cat}$\tablenotemark{*1}} & 
\colhead{$I_{\rm int}$\tablenotemark{*2}} & 
\colhead{$\sigma$} & 
\colhead{$T_{\rm mb}$} &
\colhead{$\sigma$} & 
\colhead{$V_{\rm exp}$} &
\colhead{$\sigma$} & 
\colhead{Note\tablenotemark{*3}} 
}
\startdata
U131352       &$                                $    &131352.15 &4.40 &           &  2.96 &0.37 &   90 & 45 &23.20 & 0.89    &s   \\
CCS           &$N=10-9,J=11-10                  $    &131551.97 &     &131551.97  &  0.68 &0.32 &   26 & 44 &14.50 &         &+   \\   
AlF           &$J=4-3                           $    &131895.82 &2.35 &131899.23a &  1.33 &0.12 &   30 & 21 &25.00 & 3.55    &\&  \\
\enddata
\tablenotetext{*}{The complete version of this table is in the electronic edition of the 
Astrophysical Journal Supplement Series. This is only a sample.}
\tablenotetext{*1}{Catalogue frequencies may be followed by an one
  letter coding [a, b or d]: (a) average frequency of hyperfine
  structure components; (b)the same as (a) but for MgNC, the hyperfine
  structure components of the two fine structure lines partially mix together and frequency is the average within each find structure level; (d) average frequency of the unresolved lambda doublets of SiC.}
\tablenotetext{*2}{Entries followed by a star mark are line
  intensities derived by truncated parabola fitting.}
\tablenotetext{*3}{The Note column only contains coding for line
  blending status: (s) single line; (+)line fitted with fixed
  frequency and line half width of 14.5\,km\,s$^{-1}$; (@)line too narrow;
  (*)differentiable blended lines of the same species;
  (\$)differentiable blended lines of different species;
  (\&)undifferentiable blended lines of the same species (e.g.,
  hyperfine structure or lambda doublet or some asymmetric top transitions); (\#)undifferentiable blended lines of different species.}
\end{deluxetable}

\clearpage
\begin{deluxetable}{lllllrrrrrrr}
\tabletypesize{\scriptsize}
\tabletypesize{\tiny}
\tablecaption{\small Line parameters of unidentified lines. The units are MHz for $\nu_{\rm obs}$ and $\nu_{\rm cat}$, K\,km\,s$^{-1}$ for $I_{\rm int}$, mK for $T_{\rm mb}$, and km\,s$^{-1}$ for $V_{\rm exp}$; the same for corresponding uncertainties. \label{tab5}}
\tablewidth{0pt}
\tablehead{
\colhead{Species} & 
\colhead{Trans} & 
\colhead{$\nu_{\rm obs}$} & 
\colhead{$\sigma$} &
\colhead{$\nu_{\rm cat}$\tablenotemark{*1}} & 
\colhead{$I_{\rm int}$\tablenotemark{*2}} & 
\colhead{$\sigma$} & 
\colhead{$T_{\rm mb}$} &
\colhead{$\sigma$} & 
\colhead{$V_{\rm exp}$} &
\colhead{$\sigma$} & 
\colhead{Note\tablenotemark{*3}} 
}
\startdata
U131352  &                           & 131352.15 &4.40  &           & 2.96 &0.37   &   90 & 45 &23.20 & 0.89   & s  \\
U141000  &                           & 140999.58 &4.40  &           & 0.12 &0.07   &    2 &  8 &14.50 &        & +  \\
U227219  &                           & 227219.26 &4.00  &           & 0.59*&0.11   &   20 & 25 &17.52 & 5.28   & s  \\  
U230601  &                           & 230601.00 &5.00  &           & 0.55*&0.05   &   18 & 18 & 8.79 & 6.49   & @ \\  
U234078  &                           & 234077.89 &3.00  &           & 0.46*&0.08   &   14 & 18 &12.48 & 3.84   & s  \\  
U235017  &                           & 235016.69 &2.00  &           & 0.51 &0.13   &   10 & 14 &19.20 & 1.28   & s  \\
U235175  &                           & 235174.70 &4.00  &           & 0.22 &0.28   &   22 & 10 & 4.97 & 1.27   & @ \\
U236167  &                           & 236167.09 &2.50  &           & 0.63 &0.21   &   26 & 16 &15.00 & 1.27   & s  \\
\enddata
\tablenotetext{*1}{The same as *1 in Table~\ref{tab4}.}
\tablenotetext{*2}{The same as *2 in Table~\ref{tab4}.}
\tablenotetext{*3}{The same as *3 in Table~\ref{tab4}.}
\end{deluxetable}

\clearpage
\begin{deluxetable}{lllllrrrrrrr}
\tabletypesize{\scriptsize}
\tabletypesize{\tiny}
\tablecaption{\small Line parameters for silicon containing species. The units are MHz for $\nu_{\rm obs}$ and $\nu_{\rm cat}$, K\,km\,s$^{-1}$ for $I_{\rm int}$, mK for $T_{\rm mb}$, and km\,s$^{-1}$ for $V_{\rm exp}$; the same for corresponding uncertainties. \label{tab6}}
\tablewidth{0pt}
\tablehead{
\colhead{Species} & 
\colhead{Trans} & 
\colhead{$\nu_{\rm obs}$} & 
\colhead{$\sigma$} &
\colhead{$\nu_{\rm cat}$\tablenotemark{*1}} & 
\colhead{$I_{\rm int}$\tablenotemark{*2}} & 
\colhead{$\sigma$} & 
\colhead{$T_{\rm mb}$} &
\colhead{$\sigma$} & 
\colhead{$V_{\rm exp}$} &
\colhead{$\sigma$} & 
\colhead{Note\tablenotemark{*3}} 
}
\startdata
SiC        &$\Omega=1,\Lambda=1,f,J=6-5   $      &242968.95 &     &242968.95  &  0.55 &0.11 &   20 & 18 &14.50 &        & \&  \\  
SiC        &$\Omega=1,\Lambda=1,e,J=6-5   $      &243176.94 &     &243176.94  &  0.58*&0.10 &   30 & 17 &14.50 &        & \$  \\
SiC        &$\Omega=2,\Lambda=1,e\&f,J=4-3$      &157493.57 &0.39 &157494.09d &  2.30 &0.15 &   76 & 31 &13.20 & 0.74   & \&  \\  
SiC        &$\Omega=2,\Lambda=1,e\&f,J=6-5$      &236287.79 &1.00 &236287.61d &  2.73 &0.15 &   83 & 25 &13.70 & 1.27   & \&  \\  
SiN        &$N=6-5,J=11/2-9/2              $     &261648.90 &1.00 &261650.23a &  1.85 &0.16 &   48 & 25 &12.40 & 1.15   & \&  \\  
SiN        &$N=6-5,J=13/2-11/2             $     &262154.37 &1.00 &262155.46a &  0.89 &0.17 &   24 & 25 & 9.99 & 1.14   & \&  \\  
SiO        &$J=6-5                         $     &260517.67 &1.00 &260518.02  & 63.90 &0.10 & 2700 & 15 &12.80 & 1.15   & s   \\ 
SiO        &$v=1,J=6-5                     $     &258707.39 &     &258707.39  &  1.37 &0.25 &   37 & 43 &14.50 &        & \$  \\
\enddata
\tablenotetext{*1}{The same as *1 in Table~\ref{tab4}.}
\tablenotetext{*2}{The same as *2 in Table~\ref{tab4}.}
\tablenotetext{*3}{The same as *3 in Table~\ref{tab4}.}
\end{deluxetable}

\clearpage
\begin{deluxetable}{lllllrrrrrrr}
\tabletypesize{\scriptsize}
\tabletypesize{\tiny}
\tablecaption{\small Line parameters SiCC lines. The units are MHz for $\nu_{\rm obs}$ and $\nu_{\rm cat}$, K\,km\,s$^{-1}$ for $I_{\rm int}$, mK for $T_{\rm mb}$, and km\,s$^{-1}$ for $V_{\rm exp}$; the same for corresponding uncertainties. \label{tab7}}
\tablewidth{0pt}
\tablehead{
\colhead{Species} & 
\colhead{Trans} & 
\colhead{$\nu_{\rm obs}$} & 
\colhead{$\sigma$} &
\colhead{$\nu_{\rm cat}$\tablenotemark{*1}} & 
\colhead{$I_{\rm int}$\tablenotemark{*2}} & 
\colhead{$\sigma$} & 
\colhead{$T_{\rm mb}$} &
\colhead{$\sigma$} & 
\colhead{$V_{\rm exp}$} &
\colhead{$\sigma$} & 
\colhead{Note\tablenotemark{*3}} 
}
\startdata
SiCC          &$6_{0,6}-5_{0,5}         $             &137180.73 &0.39 &137180.79 &  24.50 &0.17 &  884 & 33 &13.90 & 0.85 &   s   \\ 
SiCC          &$6_{2,4}-5_{2,3}         $             &145325.81 &0.39 &145325.87 &  20.40 &0.12 &  716 & 24 &13.90 & 0.80 &   s   \\ 
SiCC          &$6_{2,5}-5_{2,4}         $             &140920.08 &0.39 &140920.18 &  20.20 &0.05 &  705 & 11 &13.70 & 0.83 &   s   \\ 
SiCC          &$6_{4,2}-5_{4,1}         $             &141755.25 &0.39 &141755.36 &  19.10 &0.11 &  371 & 15 &14.00 & 0.82 &   *   \\ 
SiCC          &$6_{4,3}-5_{4,2}         $             &141751.42 &0.39 &141751.49 &  19.60 &0.11 &  424 & 15 &13.90 & 0.82 &   *   \\ 
SiCC          &$7_{0,7}-6_{0,6}         $             &158499.13 &0.39 &158499.24 &  27.50 &0.15 &  916 & 32 &13.70 & 0.74 &   s   \\ 
SiCC          &$9_{2,7}-8_{2,6}         $             &222009.36 &1.00 &222009.38 &  24.10 &0.12 &  842 & 19 &14.00 & 1.35 &   s   \\ 
SiCC          &$10_{0,10}-9_{0,9}       $             &220773.68 &1.00 &220773.69 &  24.60 &0.14 &  984 & 21 &13.70 & 1.36 &   s   \\
\enddata
\tablenotetext{*1}{The same as *1 in Table~\ref{tab4}.}
\tablenotetext{*2}{The same as *2 in Table~\ref{tab4}.}
\tablenotetext{*3}{The same as *3 in Table~\ref{tab4}.}
\end{deluxetable}

\clearpage
\begin{deluxetable}{lllllrrrrrrr}
\tabletypesize{\scriptsize}
\tabletypesize{\tiny}
\tablecaption{\small Line parameters for metal containing species. The units are MHz for $\nu_{\rm obs}$ and $\nu_{\rm cat}$, K\,km\,s$^{-1}$ for $I_{\rm int}$, mK for $T_{\rm mb}$, and km\,s$^{-1}$ for $V_{\rm exp}$; the same for corresponding uncertainties. \label{tab8}}
\tablewidth{0pt}
\tablehead{
\colhead{Species} & 
\colhead{Trans} & 
\colhead{$\nu_{\rm obs}$} & 
\colhead{$\sigma$} &
\colhead{$\nu_{\rm cat}$\tablenotemark{*1}} & 
\colhead{$I_{\rm int}$\tablenotemark{*2}} & 
\colhead{$\sigma$} & 
\colhead{$T_{\rm mb}$} &
\colhead{$\sigma$} & 
\colhead{$V_{\rm exp}$} &
\colhead{$\sigma$} & 
\colhead{Note\tablenotemark{*3}} 
}
\startdata
AlCl          &$J=10-9                  $           &145744.46 &     &145744.46a &  0.34*&0.06 &   13 & 13 &14.50 & 1.21 &  \$\& \\
AlCl          &$J=16-15                 $           &233116.28 &1.00 &233116.32a &  1.96 &0.08 &   70 & 13 &13.40 & 1.29 &   \&  \\ 
AlCl          &$J=18-17                 $           &262220.12 &     &262219.25a &  2.29*&0.22 &   58 & 25 &13.20 & 1.14 &  \&\$ \\  
Al$^{37}$Cl   &$J=11-10                 $           &156546.44 &0.39 &156547.00a &  0.55 &0.09 &   23 & 18 &13.30 & 0.75 &   \&  \\ 
Al$^{37}$Cl   &$J=16-15                 $           &227643.31 &     &227643.31a &  0.46 &0.12 &   25 & 19 &14.50 &      &   \&  \\ 
Al$^{37}$Cl   &$J=17-16                 $           &241855.85 &     &241854.99a &  0.60 &0.09 &   22 & 15 &14.50 &      &  \&\# \\  
Al$^{37}$Cl   &$J=18-17                 $           &256063.28 &1.00 &256063.74a &  0.72 &0.08 &   24 & 14 &13.50 & 1.17 &   \&  \\ 
AlF           &$J=4-3                   $           &131895.03 &2.35 &131899.23a &  1.33 &0.12 &   30 & 21 &25.00 & 3.55 &   \&  \\
\enddata
\tablenotetext{*1}{The same as *1 in Table~\ref{tab4}.}
\tablenotetext{*2}{The same as *2 in Table~\ref{tab4}.}
\tablenotetext{*3}{The same as *3 in Table~\ref{tab4}.}
\end{deluxetable}

\clearpage
\begin{deluxetable}{lllllrrrrrrr}
\tabletypesize{\scriptsize}
\tabletypesize{\tiny}
\tablecaption{\small Line parameters for hydrocarbon compounds. The units are MHz for $\nu_{\rm obs}$ and $\nu_{\rm cat}$, K\,km\,s$^{-1}$ for $I_{\rm int}$, mK for $T_{\rm mb}$, and km\,s$^{-1}$ for $V_{\rm exp}$; the same for corresponding uncertainties. \label{tab9}}
\tablewidth{0pt}
\tablehead{
\colhead{Species} & 
\colhead{Trans} & 
\colhead{$\nu_{\rm obs}$} & 
\colhead{$\sigma$} &
\colhead{$\nu_{\rm cat}$\tablenotemark{*1}} & 
\colhead{$I_{\rm int}$\tablenotemark{*2}} & 
\colhead{$\sigma$} & 
\colhead{$T_{\rm mb}$} &
\colhead{$\sigma$} & 
\colhead{$V_{\rm exp}$} &
\colhead{$\sigma$} & 
\colhead{Note\tablenotemark{*3}} 
}
\startdata
CCH                &$(N,J,F)=(3,\frac{5}{2},3)-(2,\frac{5}{2},3)                             $  &262208.59 &1.00 &262208.44 &   5.33*&0.25 & 129  & 25 &13.80 & 1.14 &   \$      \\
CCH                &$(N,J,F)=(3,\frac{5}{2},2)-(2,\frac{3}{2},2)                             $  &262079.60 &1.00 &262078.78 &   6.54*&0.57 & 226  & 25 &13.80 & 1.14 &   *       \\
CCH                &$(N,J,F)=(3,\frac{5}{2},2)-(2,\frac{3}{2},1)                             $  &262065.62 &1.00 &262067.33 &  69.60*&0.75 &2390  & 25 &13.90 & 1.14 &  *\&      \\
CCH                &$(N,J,F)=(3,\frac{5}{2},3)-(2,\frac{3}{2},2)                             $  &          &     &262064.84 &        &     &      &    &      &      &  *\&      \\
CCH                &$(N,J)=(3,\frac{7}{2})-(2,\frac{5}{2})                                   $  &262004.73 &1.00 &262005.32a&  96.20*&1.08 &3520  & 25 &13.90 & 1.14 &  \$\&     \\
$^{13}$CCH         &$(N,J,F_1,F)=(3,\frac{5}{2},2,\frac{5}{2})-(2,\frac{3}{2},1,\frac{3}{2}) $  &252480.92 &     &252480.92 &   1.39*&0.25 &  35  &43  &14.50 &      &   *       \\
$^{13}$CCH         &$(N,J,F_1,F)=(3,\frac{5}{2},3,\frac{5}{2})-(2,\frac{3}{2},2,\frac{3}{2}) $  &252468.78 &     &252468.78 &   0.02*&0.02 &   2  &44  &14.50 &      &   *       \\
$^{13}$CCH         &$(N,J,F_1,F)=(3,\frac{5}{2},3,\frac{7}{2})-(2,\frac{3}{2},2,\frac{5}{2}) $  &252457.86 &     &252457.86 &   0.74*&0.29 &  19  &43  &14.50 &      &   *       \\
\enddata
\tablenotetext{*1}{The same as *1 in Table~\ref{tab4}.}
\tablenotetext{*2}{The same as *2 in Table~\ref{tab4}.}
\tablenotetext{*3}{The same as *3 in Table~\ref{tab4}.}
\end{deluxetable}

\clearpage
\begin{deluxetable}{lllllrrrrrrr}
\tabletypesize{\scriptsize}
\tabletypesize{\tiny}
\tablecaption{\small Line parameters for nitrogen containing species. The units are MHz for $\nu_{\rm obs}$ and $\nu_{\rm cat}$, K\,km\,s$^{-1}$ for $I_{\rm int}$, mK for $T_{\rm mb}$, and km\,s$^{-1}$ for $V_{\rm exp}$; the same for corresponding uncertainties. \label{tab10}}
\tablewidth{0pt}
\tablehead{
\colhead{Species} & 
\colhead{Trans} & 
\colhead{$\nu_{\rm obs}$} & 
\colhead{$\sigma$} &
\colhead{$\nu_{\rm cat}$\tablenotemark{*1}} & 
\colhead{$I_{\rm int}$\tablenotemark{*2}} & 
\colhead{$\sigma$} & 
\colhead{$T_{\rm mb}$} &
\colhead{$\sigma$} & 
\colhead{$V_{\rm exp}$} &
\colhead{$\sigma$} & 
\colhead{Note\tablenotemark{*3}} 
}
\startdata
CN            &$(N,J,F)=(2,\frac{3}{2},\frac{1}{2})-(1,\frac{3}{2},\frac{1}{2})   $      &226286.81 &1.00 &226287.42 &   2.55*&0.13 &   78 & 17 &14.60 & 1.32 &  *\&    \\
CN            &$(N,J,F)=(2,\frac{3}{2},\frac{1}{2})-(1,\frac{3}{2},\frac{3}{2})   $      &226301.21 &1.00 &226298.92 &   2.37*&0.16 &   65 & 17 &16.40 & 1.32 &  *\&    \\
CN            &$(N,J,F)=(2,\frac{3}{2},\frac{3}{2})-(1,\frac{3}{2},\frac{3}{2})   $      &226314.84 &1.00 &226314.55 &   3.26*&0.14 &  101 & 17 &14.40 & 1.32 &  *\&    \\
CN            &$(N,J,F)=(2,\frac{3}{2},\frac{3}{2})-(1,\frac{3}{2},\frac{1}{2})   $      &          &     &226303.08 &        &     &      &    &      &      &  *\&    \\
CN            &$(N,J,F)=(2,\frac{3}{2},\frac{3}{2})-(1,\frac{3}{2},\frac{5}{2})   $      &226341.45 &1.00 &226332.53 &   3.52*&0.12 &  113 & 17 &13.90 & 1.32 &  *\&    \\
CN            &$(N,J,F)=(2,\frac{3}{2},\frac{5}{2})-(1,\frac{3}{2},\frac{5}{2})   $      &226359.79 &1.00 &226359.87 &   6.43*&0.12 &  213 & 17 &13.50 & 1.32 &  *\&    \\
CN            &$(N,J,F)=(2,\frac{3}{2},\frac{5}{2})-(1,\frac{3}{2},\frac{3}{2})   $      &          &     &226341.94 &        &     &      &    &      &      &  *\&    \\
CN            &$(N,J,F)=(2,\frac{3}{2},\frac{1}{2})-(1,\frac{1}{2},\frac{1}{2})   $      &226661.71 &1.00 &226663.70 &   7.29*&0.06 &  267 & 19 &13.10 & 1.32 &   *     \\
\enddata
\tablenotetext{.}{(to be continued)}
\end{deluxetable}

\clearpage
\begin{deluxetable}{lllllrrrrrrr}
\tabletypesize{\scriptsize}
\tabletypesize{\tiny}
\tablecaption{\small Line parameters for oxygen and sulfur containing species. The units are MHz for $\nu_{\rm obs}$ and $\nu_{\rm cat}$, K\,km\,s$^{-1}$ for $I_{\rm int}$, mK for $T_{\rm mb}$, and km\,s$^{-1}$ for $V_{\rm exp}$; the same for corresponding uncertainties. \label{tab11}}
\tablewidth{0pt}
\tablehead{
\colhead{Species} & 
\colhead{Trans} & 
\colhead{$\nu_{\rm obs}$} & 
\colhead{$\sigma$} &
\colhead{$\nu_{\rm cat}$\tablenotemark{*1}} & 
\colhead{$I_{\rm int}$\tablenotemark{*2}} & 
\colhead{$\sigma$} & 
\colhead{$T_{\rm mb}$} &
\colhead{$\sigma$} & 
\colhead{$V_{\rm exp}$} &
\colhead{$\sigma$} & 
\colhead{Note\tablenotemark{*3}} 
}
\startdata
CO               &$J=2-1                     $         &230537.32 &1.00 &230538.00 & 561.00 &0.10 &26600 & 15 &13.60 & 1.30 &   s   \\ 
$^{13}$CO        &$J=2-1                     $         &220398.48 &1.00 &220398.67 &  56.30 &0.09 & 1650 & 14 &14.20 & 1.36 &   s   \\ 
C$^{17}$O        &$J=2-1                     $         &224714.37 &1.00 &224714.39 &   3.45 &0.11 &   78 & 18 &14.20 & 1.33 &   s   \\ 
C$^{18}$O        &$J=2-1                     $         &219560.23 &1.00 &219560.36 &   2.94 &0.09 &   64 & 14 &14.40 & 1.37 &   s   \\ 
H$_2$CO          &$2_{1,1}-1_{1,0}           $         &150498.33 &     &150498.33 &   0.15 &0.14 &    3 & 18 &14.50 &      &   +   \\ 
H$_2$CO          &$2_{1,2}-1_{1,1}           $         &140839.50 &     &140839.50 &   0.22 &0.07 &   21 & 10 &14.50 &      &   +   \\
CS               &$J=3-2                     $         &146968.79 &0.39 &146969.03 & 113.00 &0.08 & 4900 & 16 &13.00 & 0.80 &   s   \\ 
CS               &$J=5-4                     $         &244934.91 &1.00 &244935.64 & 220.00 &0.30 & 9910 & 47 &12.50 & 1.22 &   s   \\
\enddata
\tablenotetext{*1}{The same as *1 in Table~\ref{tab4}.}
\tablenotetext{*2}{The same as *2 in Table~\ref{tab4}.}
\tablenotetext{*3}{The same as *3 in Table~\ref{tab4}.}
\end{deluxetable}

\clearpage
\begin{deluxetable}{rclcllccccl}
\tabletypesize{\scriptsize}
\tablecaption{Column density and excitation temperature of extended species and some results from the literature. The unit is $10^{14}{\rm cm}^{-2}$ for column density and K for temperature. \label{coldens}}
\tablewidth{0pt}
\tablehead{
\colhead{Mol}                          & 
\colhead{$T_{\rm ex}$}                 & 
\colhead{$\sigma\tablenotemark{\dag}$} & 
\colhead{$N_{\rm col}$}                & 
\colhead{$\sigma\tablenotemark{\dag}$} &
\colhead{$\theta_s$}                   & 
\colhead{Lit.$_{\theta}$}              & 
\colhead{$T_{\rm ex,lit}$}             & 
\colhead{$N_{\rm lit}$}                & 
\colhead{Lit.$_{T,N}$}                 & 
\colhead{Notes on our results} 
}
\startdata
C$_3$H         &25  &$  5  $     & 1.32     &$  0.35 $     &$30\arcsec$    &1    & 20    & 0.7      &7   & \\
               &    &            &          &              &               &     & 8.5   &  0.56    &8   & \\
C$_4$H         &53  &$  3  $     &81        &$  11   $     &$30\arcsec$    &1    & 35    & 30       &7   & ($v=0$) \\
               &    &            &          &              &               &     & 48    & 56       &9   & \\
               &    &            &          &              &               &     & 15    & 24       &8   & \\
C$_4$H         &53  &$  4  $     &59        &$  21   $     &$30\arcsec$    &2    &       &          &    & ($\nu_7=1$  ) \\
C$_4$H         &51  &$  3  $     & 77       &$  65   $     &$30\arcsec$    &2    &       &          &    & ($\nu_7=2^0$) \\
{\it c}-C$_3$H$_2$   &8   &$  6  $     & 1.22     &$  2.03 $     &$30\arcsec$    &2    & 20    & 0.16     &8   & (para )       \\
\enddata
\tablenotetext{\dag}{Results without uncertainty are derived by two-point fitting in 
the rotation diagram. Note that the $1\sigma$ uncertainty of $T_{\rm ex}$ of C$_3$H, {\it c}-C$_3$H$_2$, C$_3$N, H$^{13}$CCCN, HC$^{13}$CCN, HCC$^{13}$CN, SiCC (all 4 entries) and MgNC only includes the $10\%$ relative calibration uncertainty because they only have reliable line intensities from one of our two telescopes, while that of the other entries and the uncertainty of all $N_{\rm col}$ values include the $20\%$ absolute calibration uncertainty.}
\tablerefs{
(1) Gu{\'e}lin et al. 1993;
(2) assumed to the same as its main isotopomer;
(3) Bieging \& Taffala 1993;
(4) Gu{\'e}lin et al. 1996;
(5) Cernicharo et al. 1989;
(6) Lucas et al. 1995;
(7) Cernicharo et al. 2000;
(8) Kawaguchi et al. 1995;
(9) very et al. 1992;
(10) Gu{\'e}lin et al. 1991;
(11) Groesbeck et al. 1994;
(12) Thaddeus et al. 1984.}
\end{deluxetable}

\clearpage
\begin{table*}[tH]
\begin{center}
\caption{Isotopic ratios determined from intensity ratios of optically thin lines for 
the circumstellar envelope of IRC +10216. The $1\sigma$ uncertainty 
includes the relative calibration uncertainty of $10\%$. \label{isotope}}
\begin{tabular}{crlclcl}
\tableline\tableline
Ratios & \multicolumn{2}{l}{Our results} & \multicolumn{2}{l}{Literature}  & \multicolumn{1}{l}{Solar} & Our method\\
\tableline
12C/13C    &34.7\tablenotemark{\dag} &$\pm 4.3\tablenotemark{\dag} $ &47    &(+6,-5)      \tablenotemark{a} &   89       & SiCC                       \\
17O/18O    &1.14 &$\pm 0.17$ &1.47  &(+0.33,-0.25)\tablenotemark{b} &    0.18    & CO                         \\
16O/17O    &967  &$\pm 157 $ &840   &(+230,-170)  \tablenotemark{b} & 2695       & [13CO/C17O]*[12C/13C]      \\
16O/18O    &1172 &$\pm 348 $ &1260  &(+315,-240)  \tablenotemark{b} &  490       & [13CO/C18O]*[12C/13C]      \\
32S/33S    &93   &$\pm 10  $ &100   &(+23,-16)    \tablenotemark{a} &  125       & [13CS/C33S]*[12C/13C],     \\
           &     &           &      &                               &            & [29SiS/Si33S]*[28Si/29Si], \\
           &     &           &      &                               &            & [30SiS/Si33S]*[28Si/30Si]  \\
32S/34S    &18.9 &$\pm 1.3 $ &20.2  &(+2.6-2.1)   \tablenotemark{a} &   22.5     & [29SiS/Si34S]*[28Si/29Si], \\
           &     &           &      &                               &            & [30SiS/Si34S]*[28Si/30Si]  \\
34S/33S    &5.17 &$\pm 0.44$ &5.7   &(+0.8,-0.6)  \tablenotemark{a} &    5.5     & CS                         \\
28Si/29Si  &17.2 &$\pm 1.1 $ &18.7  &(+1.3,-1.0)  \tablenotemark{a} &   19.6     & SiCC                       \\
28Si/30Si  &24.7 &$\pm 1.8 $ &28.6  &(+1.8,-1.5)  \tablenotemark{a} &   29.8     & SiCC                       \\
29Si/30Si  &1.46 &$\pm 0.11$ &1.47  &(+0.11,-0.09)\tablenotemark{a} &    1.52    & SiS,SiCC                   \\
\tableline
\end{tabular}
%
\tablenotetext{a}{Kahane et al. 1988}
\tablenotetext{b}{Kahane et al. 1992}
\tablenotetext{\dag}{The uncertainty does not take into account line saturation effect and the value could be underestimated.}
\end{center}
\end{table*}

\clearpage
\begin{table}[tH]
\scriptsize
\begin{center}
\caption{List of narrow lines. Frequency is in MHz, $T_{\rm mb}$ is in
  mK, FWHM is in km\,s$^{-1}$ and observation date is in Y/M/D format. \label{narrowlines}}
\begin{tabular}{llllll}
\tableline\tableline
Transition                 & Freq & $T_{\rm mb}^{\rm peak}$ & FWHM  & Obs.date  & Notes\\
\tableline
HCN,$\nu_2=3^{1e},J=3-2$   & 266540.82  & $96\,\pm5$         & 7        & 2007/01/07  &      \\
SiS,$v=0,J=14-13$          & 254102.82  & $9070\,\pm 11$     &          & 2007/01/08  & blue peak on the top of thermal line     \\
SiS,$v=1,J=13-12$          & 234813.22  & $92\,\pm 17$       & 8.2      & 2005/12/15  &      \\
SiS,$v=1,J=14-13$          & 252866.47  & $295\,\pm 15$      & 10.2     & 2007/01/23  & blended with {\it c}-C$_3$H, $5_{1,4}-4_{1,3},9/2-7/2$    \\
                           &            &                    &          &             & and Si$^{13}$CC, $(11_{7,5}-10_{7,4}) \& (11_{7,4}-10_{7,3})$ \\
\tableline
\end{tabular}
\end{center}
\end{table}

\clearpage

\clearpage

\clearpage
\thispagestyle{empty}
\setlength{\voffset}{-15mm}
\begin{figure*}[tH]
\centering
\includegraphics[scale=0.78]{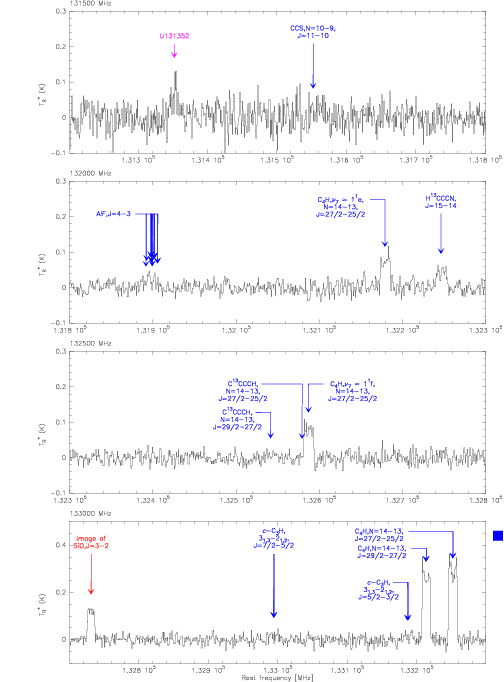}
\caption{Spectra of IRC +10216 in the frequency ranges: 131.2-160.3,
  219.5-245.5 and 251.5-267.5\,GHz. $T^*_{\rm R}$ and $T^*_{\rm A}$ are plotted for KP12M ($<200$\,GHz) and SMT ($>200$\,GHz) data respectively. Resolution is 1\,MHz in all plots. Temperature is usually plotted in -0.1 to 0.3\,K range, with exceptional plots marked by a blue square to the upper-right corner. Vertical arrows mark archival frequencies. (In an online color version, blue, magenta and red labels and arrows are used for identified, unidentified and image contamination lines, respectively, while zoom-in plots are placed in red boxes.) }
\label{spplot}
\end{figure*}
\clearpage
\setlength{\voffset}{0mm}
\centerline{\includegraphics[scale=0.78]{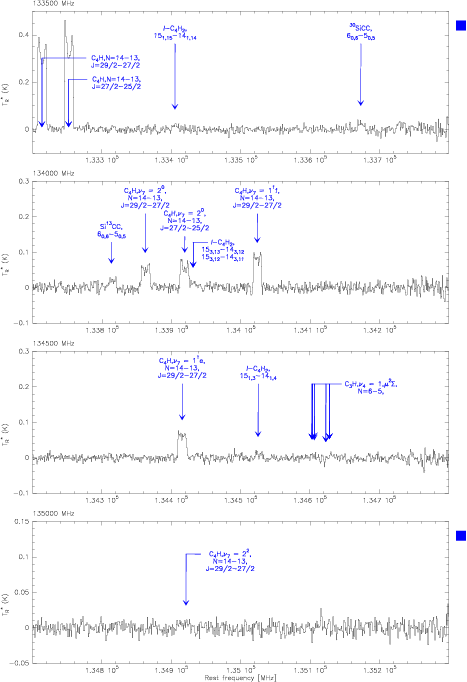}}
\centerline{Fig. \ref{spplot}. --- Continued.}
\clearpage
\centerline{\includegraphics[scale=0.78]{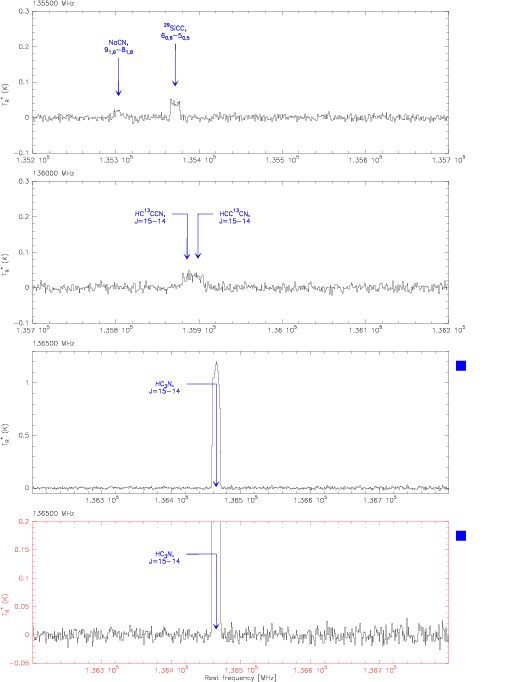}}
\centerline{Fig. \ref{spplot}. --- Continued.}
\clearpage
\centerline{\includegraphics[scale=0.78]{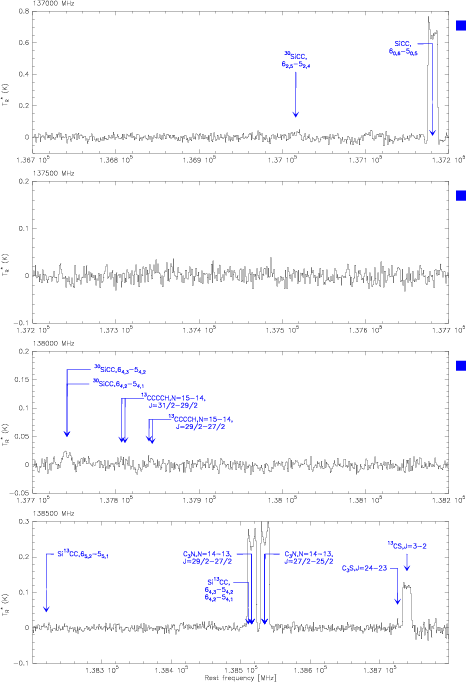}}
\centerline{Fig. \ref{spplot}. --- Continued.}
\clearpage
\centerline{\includegraphics[scale=0.78]{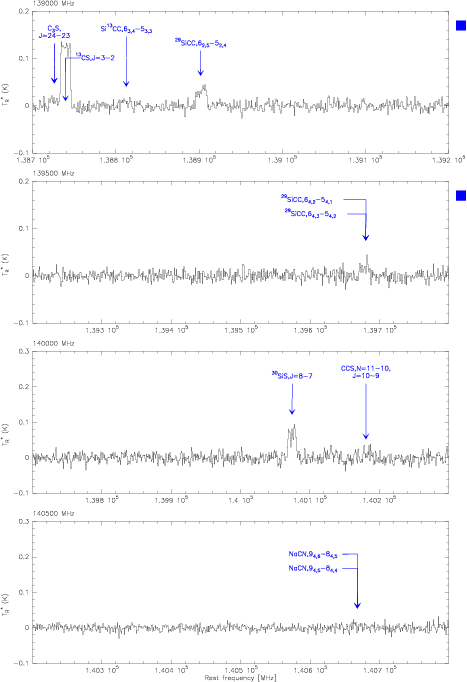}}
\centerline{Fig. \ref{spplot}. --- Continued.}
\clearpage
\centerline{\includegraphics[scale=0.78]{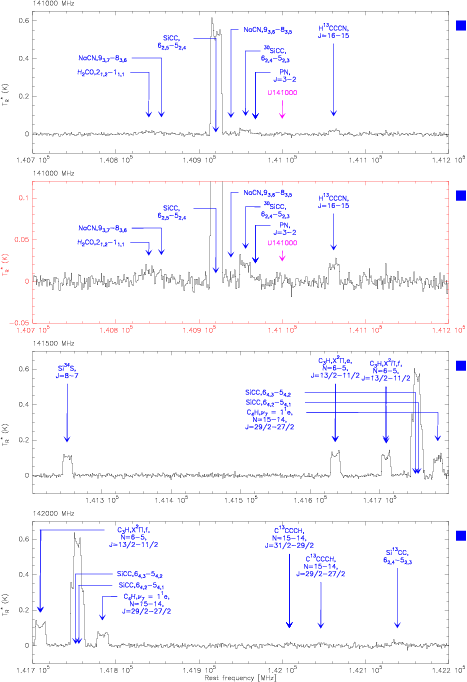}}
\centerline{Fig. \ref{spplot}. --- Continued.}
\clearpage
\centerline{\includegraphics[scale=0.78]{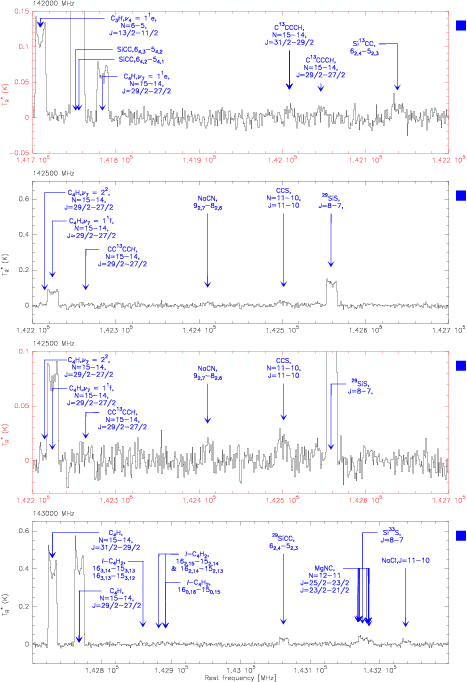}}
\centerline{Fig. \ref{spplot}. --- Continued.}
\clearpage
\centerline{\includegraphics[scale=0.78]{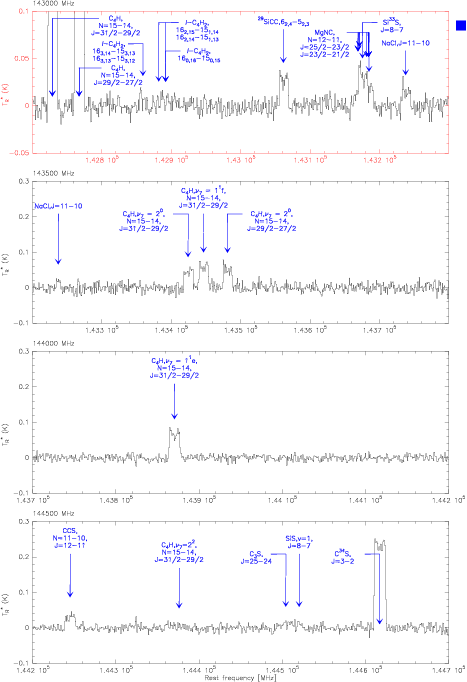}}
\centerline{Fig. \ref{spplot}. --- Continued.}
\clearpage
\centerline{\includegraphics[scale=0.78]{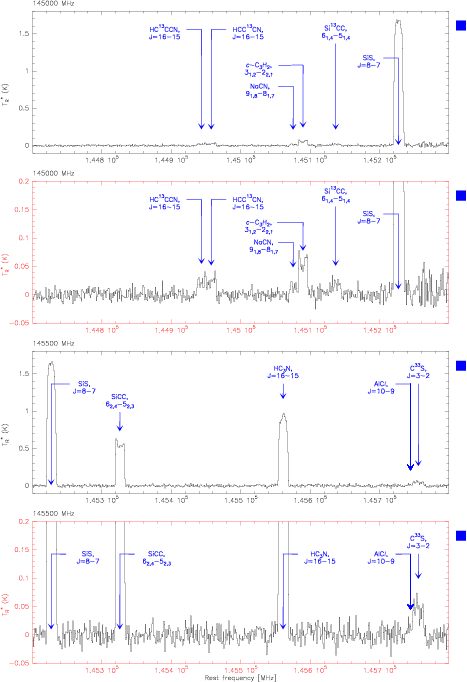}}
\centerline{Fig. \ref{spplot}. --- Continued.}
\clearpage
\centerline{\includegraphics[scale=0.78]{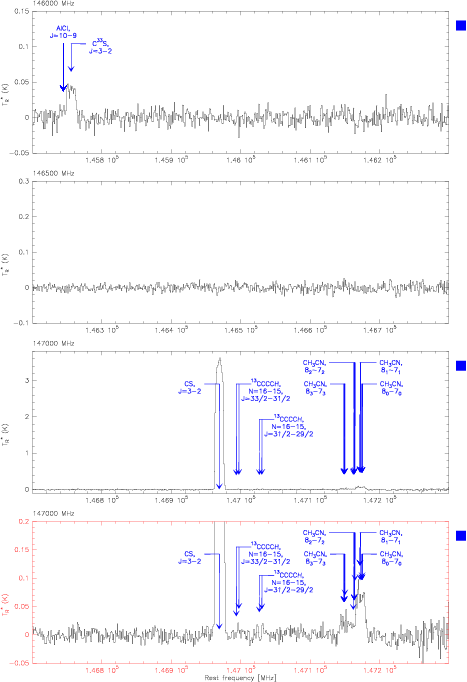}}
\centerline{Fig. \ref{spplot}. --- Continued.}
\clearpage
\centerline{\includegraphics[scale=0.78]{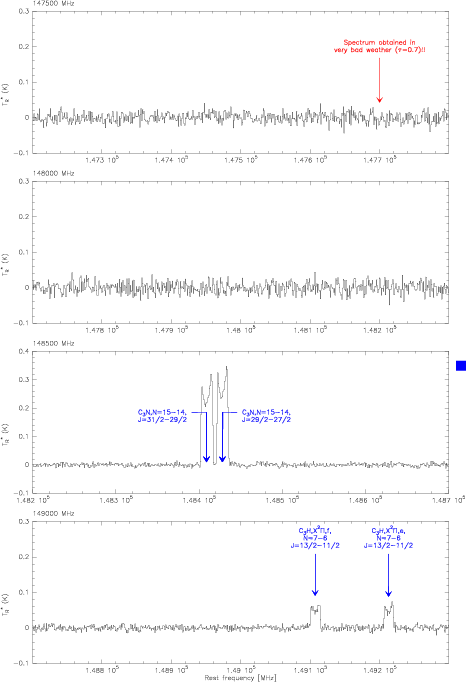}}
\centerline{Fig. \ref{spplot}. --- Continued.}
\clearpage
\centerline{\includegraphics[scale=0.78]{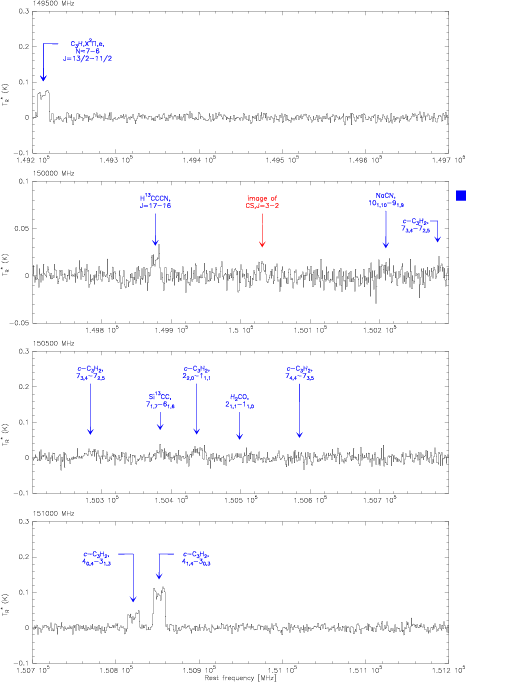}}
\centerline{Fig. \ref{spplot}. --- Continued.}
\clearpage
\centerline{\includegraphics[scale=0.78]{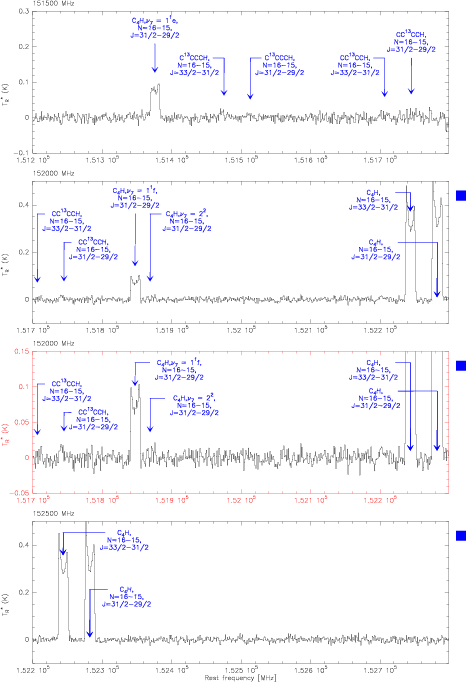}}
\centerline{Fig. \ref{spplot}. --- Continued.}
\clearpage
\centerline{\includegraphics[scale=0.78]{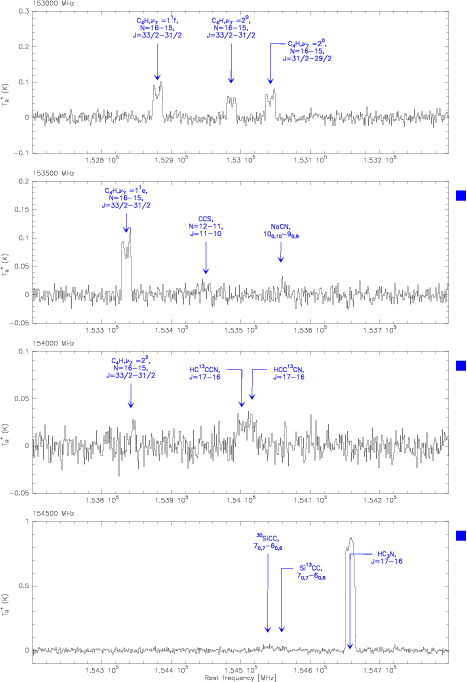}}
\centerline{Fig. \ref{spplot}. --- Continued.}
\clearpage
\centerline{\includegraphics[scale=0.78]{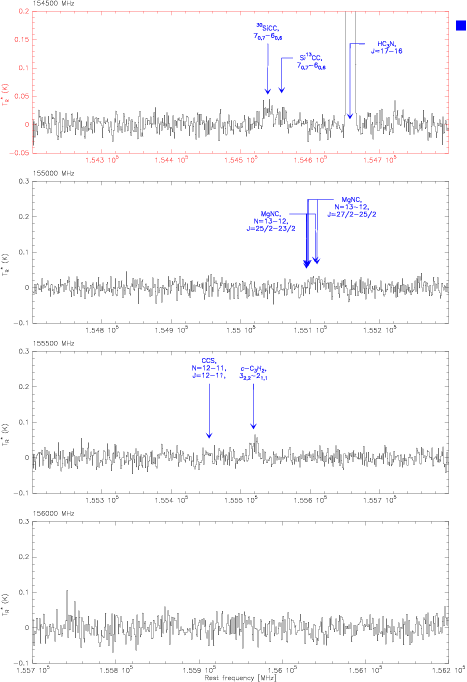}}
\centerline{Fig. \ref{spplot}. --- Continued.}
\clearpage
\centerline{\includegraphics[scale=0.78]{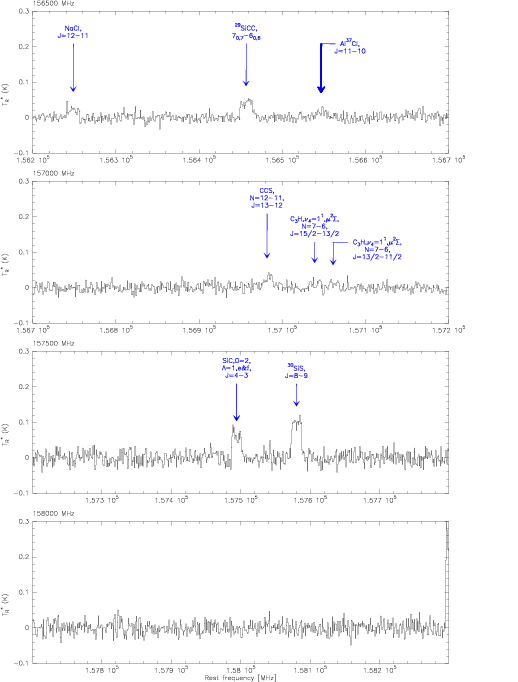}}
\centerline{Fig. \ref{spplot}. --- Continued.}
\clearpage
\centerline{\includegraphics[scale=0.78]{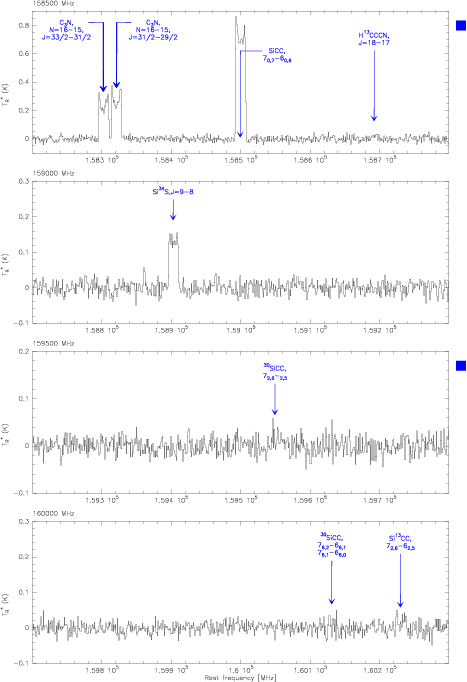}}
\centerline{Fig. \ref{spplot}. --- Continued.}
\clearpage
\centerline{\includegraphics[scale=0.78]{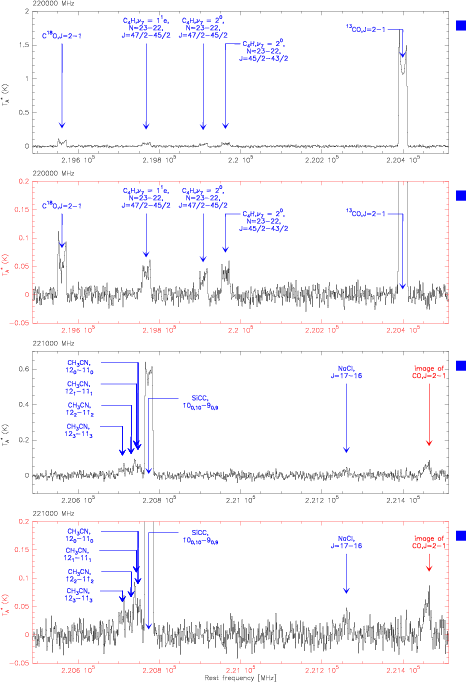}}
\centerline{Fig. \ref{spplot}. --- Continued.}
\clearpage
\centerline{\includegraphics[scale=0.78]{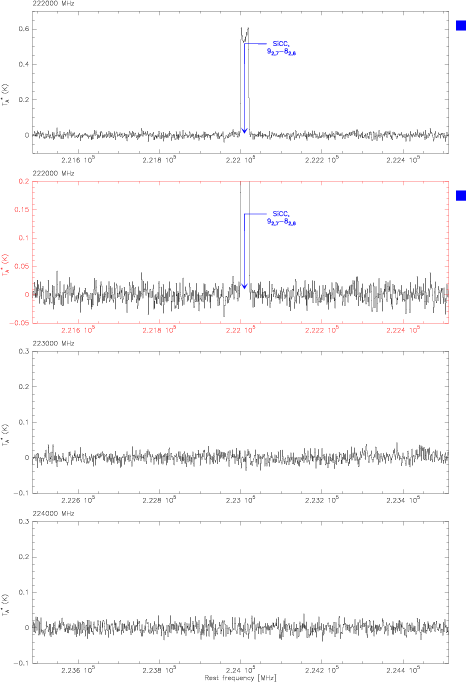}}
\centerline{Fig. \ref{spplot}. --- Continued.}
\clearpage
\centerline{\includegraphics[scale=0.78]{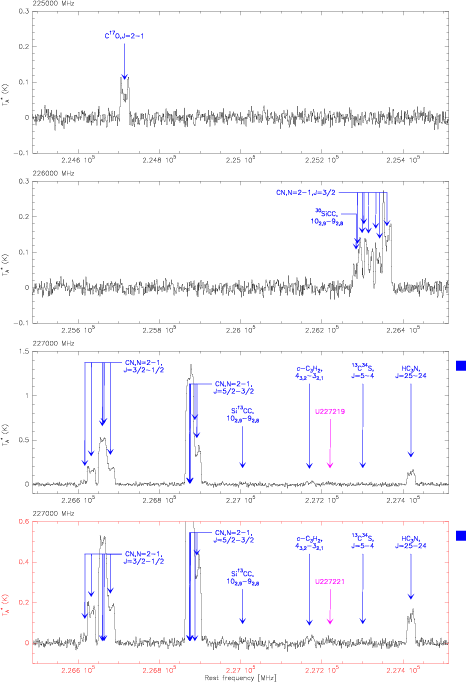}}
\centerline{Fig. \ref{spplot}. --- Continued.}
\clearpage
\centerline{\includegraphics[scale=0.78]{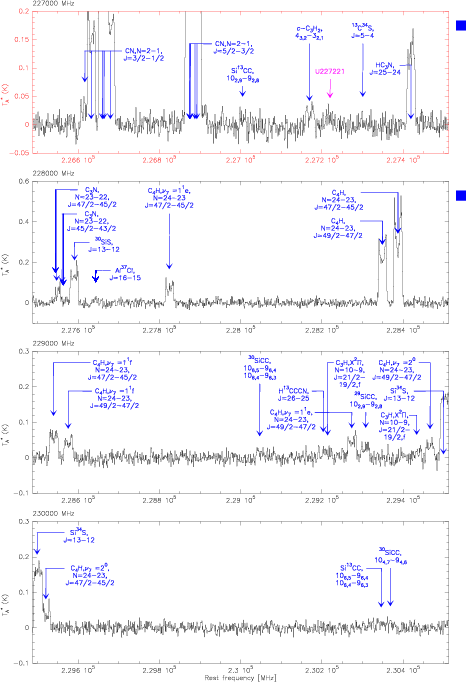}}
\centerline{Fig. \ref{spplot}. --- Continued.}
\clearpage
\centerline{\includegraphics[scale=0.78]{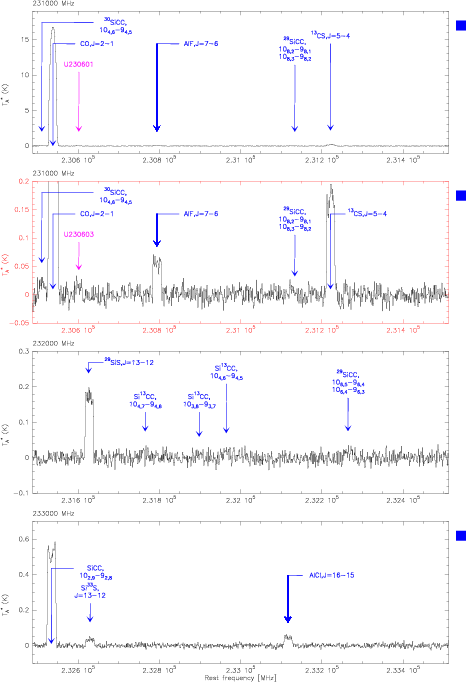}}
\centerline{Fig. \ref{spplot}. --- Continued.}
\clearpage
\centerline{\includegraphics[scale=0.78]{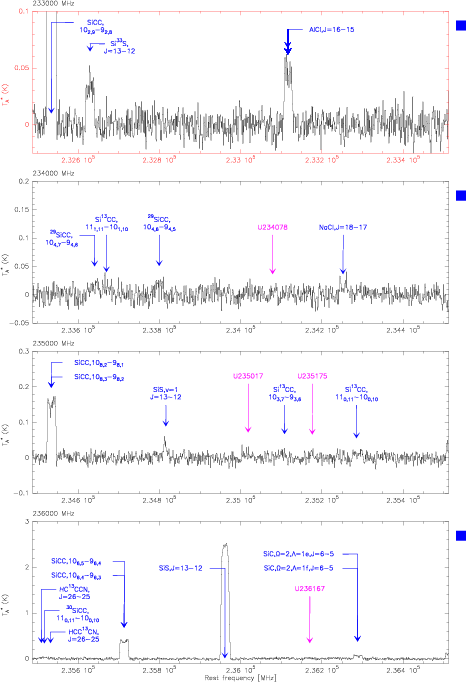}}
\centerline{Fig. \ref{spplot}. --- Continued.}
\clearpage
\centerline{\includegraphics[scale=0.78]{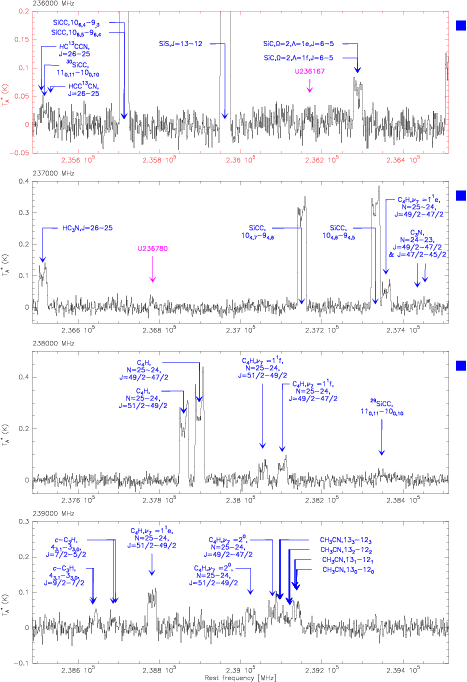}}
\centerline{Fig. \ref{spplot}. --- Continued.}
\clearpage
\centerline{\includegraphics[scale=0.78]{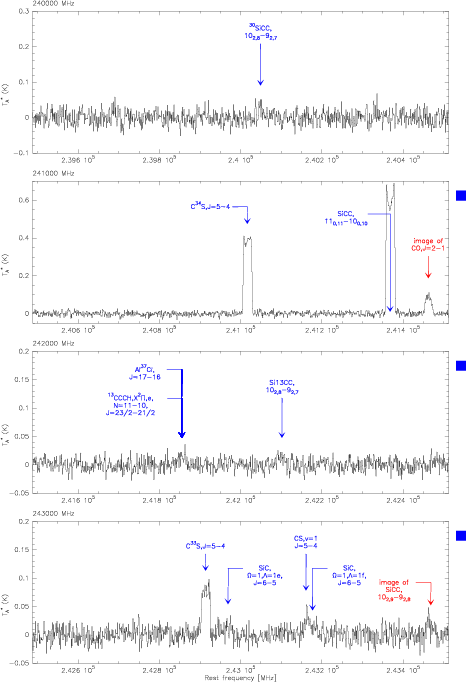}}
\centerline{Fig. \ref{spplot}. --- Continued.}
\clearpage
\centerline{\includegraphics[scale=0.78]{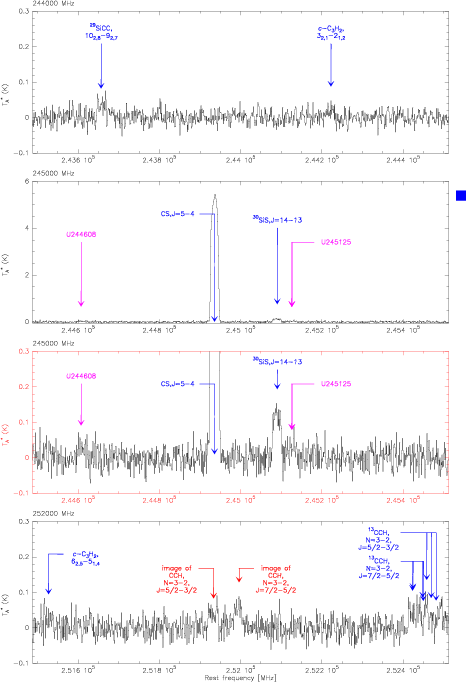}}
\centerline{Fig. \ref{spplot}. --- Continued.}
\clearpage
\centerline{\includegraphics[scale=0.78]{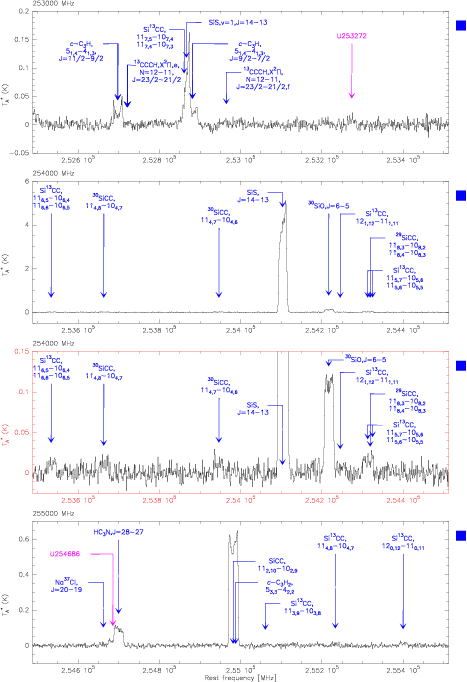}}
\centerline{Fig. \ref{spplot}. --- Continued.}
\clearpage
\centerline{\includegraphics[scale=0.78]{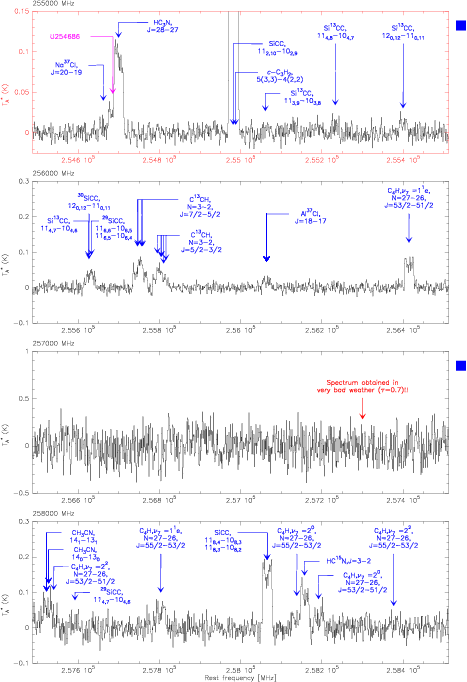}}
\centerline{Fig. \ref{spplot}. --- Continued.}
\clearpage
\centerline{\includegraphics[scale=0.78]{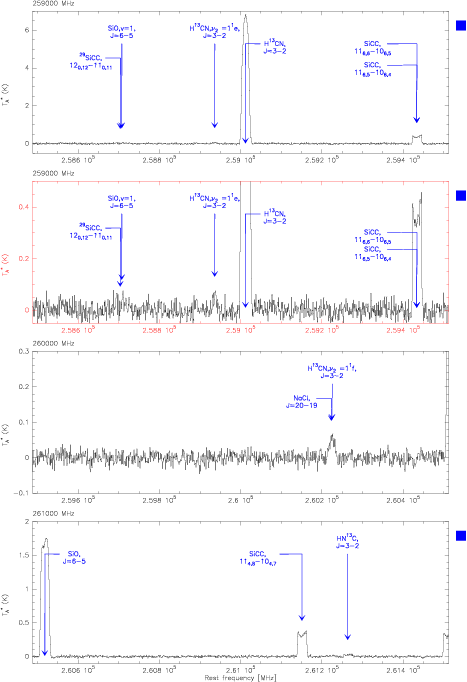}}
\centerline{Fig. \ref{spplot}. --- Continued.}
\clearpage
\centerline{\includegraphics[scale=0.78]{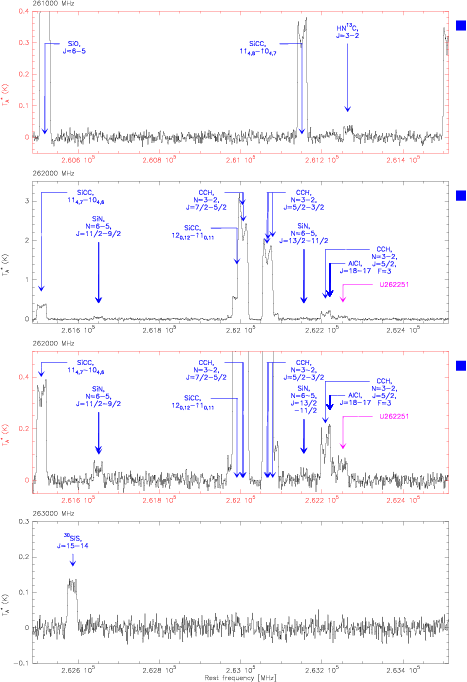}}
\centerline{Fig. \ref{spplot}. --- Continued.}
\clearpage
\centerline{\includegraphics[scale=0.78]{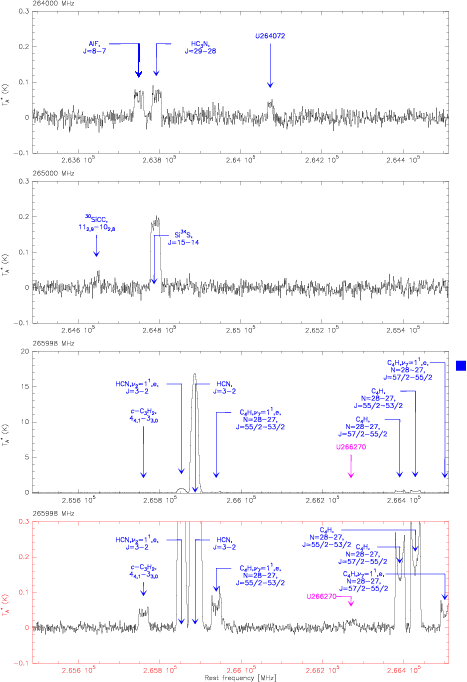}}
\centerline{Fig. \ref{spplot}. --- Continued.}
\clearpage
\centerline{\includegraphics[scale=0.78]{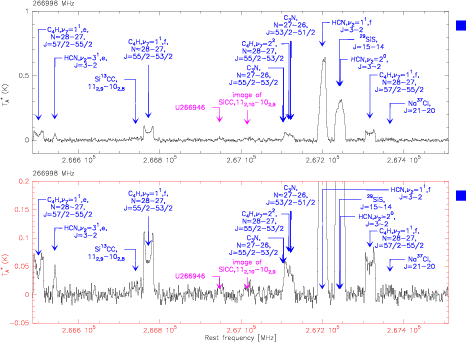}}
\centerline{Fig. \ref{spplot}. --- Continued.}

\clearpage

\end{document}